\documentclass[%
 prl,
 superscriptaddress,
reprint,
amsmath,amssymb,
floatfix,
]{revtex4-1}

\usepackage{graphicx}
\usepackage{dcolumn}
\usepackage{bm}
\usepackage{xparse}
\usepackage{hyperref}
\usepackage{siunitx}
\usepackage{bbold}
\usepackage{adjustbox}
\usepackage{stmaryrd}
\usepackage{bbm}
\usepackage{ulem}
\usepackage{color}
\DeclareSIUnit{\amagat}{amg}
\DeclareSIUnit{\sample}{Sa}

\definecolor{mygreen}{rgb}{0,0.5,0}
\definecolor{mygrey}{rgb}{0.5,0.5,0.5}
\definecolor{myred}{rgb}{0.75,0,0}
\definecolor{myblue}{rgb}{0,0,0.75}
\definecolor{mymagenta}{cmyk}{0,1,0,0.12}
\definecolor{mycyan}{cmyk}{1,0,0,0.12}
\definecolor{myorange}{rgb}{1,0.5,0}
\definecolor{myviolet}{rgb}{0.5,0.0,0.75}
\definecolor{mybrown}{cmyk}{0,0.50,1,0.41}
\definecolor{mygrey}{rgb}{0.5,0.5,0.5}

\newcommand{\grtext}[1]{{\color{mygrey}{#1}}}
\renewcommand{\grtext}[1]{{\color{mygrey}{}}}

\newcommand{\BE}{\begin{equation}}
\newcommand{\EE}{\end{equation}}
\newcommand{\BEw}{\begin{widetext}\begin{equation}}
\newcommand{\EEw}{\end{equation}\end{widetext}}

\newcommand{\BEA}{\begin{eqnarray}}
\newcommand{\EEA}{\end{eqnarray}}
\newcommand{\NAtoms}{N_\mathrm{at}}
\DeclareMathOperator{\var}{var}
\newcommand{\Rsd}{R_\mathrm{sd}}
\newcommand{\Rse}{R_\mathrm{se}}
\newcommand{\Soof}{S_{1/f}}

\begin{document}

\preprint{APS/123-QED}
\newcommand{\thetitle}{   
Anomalous noise spectra in a spin-exchange-relaxation-free alkali-metal vapor
}
\title{\thetitle}

\newcommand{\myaffiliation}{\affiliation}

\newcommand{\ICFO}
{\myaffiliation{ICFO - Institut de Ci\`encies Fot\`oniques, The Barcelona Institute of Science and Technology, 08860 Castelldefels (Barcelona), Spain}}

\newcommand{\ICREA}{\myaffiliation{ICREA - Instituci\'{o} Catalana de Recerca i Estudis Avan{\c{c}}ats, 08010 Barcelona, Spain}}

\newcommand{\HDU}
{\myaffiliation{Department of Physics, Hangzhou Dianzi University, Hangzhou 310018, China}}

\newcommand{\FLine}
{\myaffiliation{FieldLine Industries, Boulder, Colorado 80303, USA}}

\author{K. Mouloudakis}
\email{KM and JK contributed equally to this work.}
\ICFO

\author{J. Kong}
\email{KM and JK contributed equally to this work.}
\HDU

\author{A. Sierant}
\ICFO

\author{E. Arkin}
\HDU

\author{M. Hern\'{a}ndez Ruiz}
\ICFO

\author{R. Jim\'{e}nez-Mart\'{i}nez}
\FLine

\author{M. W. Mitchell}
\ICFO
\ICREA

\date{\today}

\begin{abstract}
{
We perform spin-noise spectroscopy  on an unpolarized \textsuperscript{87}$\rm{Rb}$ vapor in the spin-exchange-relaxation-free (SERF) regime. We observe noise spectral distributions that deviate strongly from Lorentzian models that accurately describe lower-density regimes. 
For example, at magnetic fields of $\sim \SI{1}{\micro \tesla}$ and \textsuperscript{87}$\rm{Rb}$ densities $\gtrsim \SI{1e14}{atoms\per\centi\meter\cubed}$ 
we observe an asymmetric spin-noise distribution in which the resonance line is depleted by about half its power, with the diverted power becoming a broad spectral component that could be mistaken for optical shot noise. The results are in good agreement with recent models accounting for correlations between the ground hyperfine states. We discuss implications for quantum sensing and absolute noise calibration in spin-squeezing and entanglement detection. The results suggest similarly anomalous spectra for other noise spectroscopies, when noise mechanisms are not aligned with system dynamics.
}
\end{abstract}

\maketitle

\newcommand{\acronym}{\rm SISNI}
\newcommand{\boldacronym}{\textbf{SISNI}}

Noise spectra, obtained by continuous observation of a system of interest without active excitation, provide information about the dynamics of nearly-undisturbed systems close to natural thermal equilibrium \cite{EinsteinAdP1905, AleksandrovJETP1981}. In recent years, noise spectroscopies have been developed for electronic \cite{TsePRL2014}, optomechanical \cite{YuCP2022}, biological \cite{ZhengNL2010} and especially spin systems \cite{MullerPhysicaE2010, CrookerN2004}, where the technique is known as spin-noise spectroscopy \cite{LuciveroPRA2017, SinitsynRPP2016, aleksandrovzapasskii, katsoprinakis, PhysRevA.104.063708, Squeezed-Spin-Noise, PhysRevApplied.17.L011001, PhysRevLett.95.216603, Diffusion, shah-vasilakis, vasilakis2015, JimenezMartinezPRL2018, ShahamPRA2020}. 
The interpretation of noise spectra benefits from weak thermal excitation, which ensures a linear response regime and allows calculation of equilibrium variances from  thermodynamic principles. Under ideal detection, noise spectra obey an ``area conservation rule''  (shape-independent net noise power) and a ``no-go theorem'' (vanishing cross-correlations) \cite{SinitsynRPP2016}. These allow absolute calibration of the participating spin number \cite{KoschorreckPRL2010a} and the identification of Lorentzian spectral features with collective or single-particle modes, and other feature shapes with inhomogeneous broadening or more complex dynamics \cite{SinitsynRPP2016, Meinel2022}. Optically-detected noise spectra also provide absolute calibration of photon shot noise (PSN) \cite{WolfgrammPRL2010, TroullinouPRL2021}. 

Here we report anomalous spin noise spectra that do not fit the above description.  Our spin system is a high-density alkali vapor that can be tuned from the spin-exchange (SE) relaxation regime into the spin-exchange-relaxation-free (SERF) regime \cite{SavukovRomalis}. In SERF, SE collisions and hyperfine interactions dominate the spin dynamics, leading to line narrowing of the magnetic resonances and a corresponding boost to the sensitivity \cite{PhysRevLett.31.273, PhysRevA.16.1877, Allred, Kominis2003}. The SERF effect is employed in biomagnetism detection \cite{BotoNI2017}, inertial sensors \cite{KornackPRL2005}, and tests of fundamental physics \cite{LeePRL2018}. Experiments show that SERF media support and preserve non-classical spin correlations, i.e., entanglement and spin squeezing \cite{MitchellNatureCom}. Theory suggests that SE collisions can preserve other non-classical states over long timescales \cite{PhysRevA.103.L010401}.

Using the quantum structure of alkali spin states, spin dynamics \cite{appelt,PRXQuantum.3.010305, PhysRevLett.80.3487}, and the regression theorem \cite{gardiner2009stochastic}, recent theory predicts spin-noise spectra in SERF regime \cite{PhysRevA.106.023112}. In agreement with this theory, we observe the following anomalous behaviors:  non-Lorentzian noise spectral features in a linear homogeneous system, flat noise backgrounds not due to PSN, and an apparent (but not real) violation of the area conservation law. We identify the underlying cause of these line-reshaping phenomena in a misalignment of the dynamical modes and the SE noise modes that drive them.

\begin{figure*}[t]
\centering
\includegraphics[width=1.7\columnwidth]{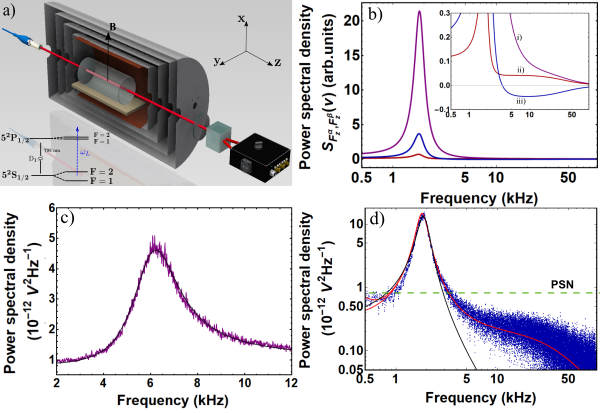}	
\caption{ Experimental setup and representative spectra. \textbf{a)} Schematic representation of the experimental setup (see text). \textbf{b)} Predicted non-Lorentzian spin-noise contributions i) $S_{\hat{F}_z^{a},\hat{F}_z^{a}}(\nu)$ , ii) $S_{\hat{F}_z^{b},\hat{F}_z^{b}}(\nu)$, iii) $S_{\hat{F}_z^{a},\hat{F}_z^{b}}(\nu)$ and  $S_{\hat{F}_z^{b},\hat{F}_z^{a}}(\nu)$, computed using Eq.~\eqref{eq:psd} and experimentally relevant parameters: $R_{\rm{se}}\approx 3.02 \times 10^5$ \si{\per\second} and $\Rsd \approx 0.03 \times 10^5 $ \si{\per\second}, corresponding to $3.4 \times 10^{14}$ \text{atoms}/\si{\centi\meter\cubed} and temperature $T=\SI{169}{\celsius}$. The magnetic field is $B=385$ \si{\nano \tesla} along the $\mathbf{\hat{x}}$ direction. \textbf{c)} Example of a non-Lorentzian spectrum at a magnetic field of $B \approx 1292$ \si{\nano\tesla} fitted to a Lorentzian plus dispersive curve (see \cite{QncuavSERFootnote} for details). \textbf{d)} Spin-noise spectra acquired at a magnetic field of $B=385$ \si{\nano\tesla} and a number density of $n \approx \SI{3.4e14}{atoms\per\centi\meter\cubed}$. The mean PSN level is depicted by the green dashed line and has been subtracted from the spectrum. Data are fitted by a Lorentzian model (black solid line) and red lines and Eq.~\eqref{eq:OpticalSpectrum} (red solid lines) with and without ``1/f-noise''. The departure from the Lorentzian spectrum is demonstrated.  
}
\label{fig:setup}
\end{figure*}

The experimental setup is shown in Fig.~\ref{fig:setup}a. Isotopically enriched $^{87}$Rb and \SI{0.12}{\amagat} of N\textsubscript{2} buffer gas are held in a cylindrical cell of \SI{12.7}{\milli\meter} diameter and \SI{30}{\milli\meter} internal length, with anti-reflection coated windows of \SI{5}{\milli\meter} thickness. The cell is placed at the center of a cylindrical, four-layer, mu-metal magnetic shield. Solenoid and shim coils are used to produce a homogeneous DC magnetic field $\mathbf{B}=(B,0,0)$ along the $\mathbf{\hat{x}}$ direction. A ceramic oven, intermittent Joule heating, and a thermocouple are used to control the cell temperature.  
An external cavity diode laser produces a linearly polarized \SI{795}{\nano\meter} beam detuned \SI{46}{\giga\hertz} to the blue of the D\textsubscript{1} line of $^{87}$Rb, monitored with a wavelength meter. The laser output, propagating along $\mathbf{\hat{z}}$, is spatially filtered with a single-mode fiber to produce a Gaussian beam with effective area  $A_{\rm{eff}}\equiv L [\int   I(x,y,z)\,dx\, dy]^2 /\int  I^2(x,y,z)\, dx\,dy\,dz \approx \SI{1.5}{\milli\meter\squared}$, where $I(x,y,z)$ is the intensity of the Gaussian beam, measured with a beam profiler, and $L$ the length of the cell \cite{Schleier-SmithPRL2010}. The effective number of atoms probed by the laser beam is $\NAtoms = n A_{\rm{eff}} L$, where $n$ is the alkali number density. Both the detuning of the light and the atomic \SI{2.4}{\giga\hertz} full-width at half maximum (FWHM) pressure-broadened optical linewidth are larger than the \SI{0.8}{\giga\hertz} hyperfine splitting of the excited state, so tensor polarizability effects are expected to be negligible \cite{PhysRev.163.12}. The transmitted light is detected by a balanced polarimeter comprised of a half-waveplate, a Wollaston prism and an amplified differential photodetector (PD). The PD signal is recorded by a \SI{24}{bit} digitizer for later processing.

The experimentally obtained noise spectra are of the form
\begin{equation}
S_\mathrm{opt}(\nu) =  S_{\rm{psn}}+ \Soof(\nu)+S_{\rm{el}}(\nu)+  S_{\rm{at}}(\nu),
\label{eq:OpticalSpectrum}
\end{equation}
where the contribution from photon shot noise (PSN) is
$S_{\rm{psn}}= 2 G^2 q_e r P \approx \SI{0.91 e-12}{\volt\squared\per\hertz}$, with $q_e \approx 1.6 \times 10^{-19}$ \si{\coulomb} being the electron charge, $r \approx 0.52$ \si{\ampere \per \watt} at \SI{795}{\nano \meter} the PD responsivity, $G =\SI{1e5}{\volt\per\ampere}$ the transimpendance gain of the PD and $P \approx 550$ \si{\micro \watt} the laser power reaching the polarimeter. $\Soof = \zeta^2 \nu^{-\beta}$, $\beta > 0$ is ``1/f noise'' with strength $\zeta^2$, and 
$S_{\rm{el}} (\nu)$ is the electronic noise of the PD and acquisition system, which in practice is about \SI{20}{\decibel} below the PSN background. The last term in Eq.~\eqref{eq:OpticalSpectrum} is the atomic spin noise spectrum, presenting a resonance feature at the spin precession frequency. 
The spin-noise power of the thermal state is a readily available noise reference, and has been used in noise calibration for spin squeezing \cite{FernholzPRL2008} and entanglement detection \cite{MitchellNatureCom} experiments.  
We note that for frequencies above $0.5$ \si{\kilo\hertz}, $\Soof(\nu)$ is negligible, thus, in the analysis that follows $S_\mathrm{opt}(\nu)$ is approximated as $S_\mathrm{opt}(\nu) \approx S_{\rm{at}}(\nu)+ S_{\rm{psn}}$.

To model the atomic spectra we employ the Ornstein-Uhlenbeck approach as derived in \cite{PhysRevA.106.023112} and further discussed in \cite{QncuavSERFootnote}. In this model, the spectra result from the stochastic dynamics of the hyperfine collective spin vectors $\hat{\mathbf{F}}^{\alpha} (t)$, $\alpha \in \{a=I+1/2, b=I-1/2 \}$, governed by
\begin{equation}
d \mathbf{\hat{X}} (t) = A \mathbf{\hat{X}}(t) dt+ Q  d{\cal{\mathbf{\hat{W}}}}(t),
\label{eq:XSDE}
\end{equation}
where $\mathbf{\hat{X}}\equiv[\hat{F}^{a}_x,\hat{F}^{a}_y,\hat{F}^{a}_z,\hat{F}^{b}_x,\hat{F}^{b}_y,\hat{F}^{b}_z]^T$, $A$ is the drift matrix, $Q$ is the noise strength matrix, and 
$d {\cal{\mathbf{\hat{W}}}}$ is a length-six vector of independent Wiener increments \cite{QncuavSERFootnote}. 
For such processes, with real $A$ and $Q$, the power spectral density matrix is \cite{gardiner2009stochastic} 
\begin{equation}
S_{\mathbf{\hat{X}},\mathbf{\hat{X}}}(\omega)  =  
 -\frac{1}{2 \pi} (A+i \omega \mathbb{1})^{-1} Q Q^T (A^T-i \omega \mathbb{1})^{-1},  \label{eq:power spec2}
\end{equation}
where $\mathbb{1}$ is the $6 \times 6$ identity matrix. In equilibrium, 
$QQ^T$ is directly related to  $A$ and to the steady-state, equal-time covariance matrix $\mathcal{R}_{\mathbf{\hat{X}},\mathbf{\hat{X}}}(0)$ by
\begin{equation}
 QQ^{T}=A \mathcal{R}_{\mathbf{\hat{X}},\mathbf{\hat{X}}}(0)+\mathcal{R}_{\mathbf{\hat{X}},\mathbf{\hat{X}}}(0) A^{T},
\label{eq:QQT}
\end{equation} 
where
\begin{equation}
\mathcal{R}_{\hat{F}_i^{\alpha},\hat{F}_j^{\beta}}(0) = \delta_{ij} \delta_{\alpha \beta} \frac{ f^{\alpha} (f^{\alpha}+1) (2f^{\alpha}+1)}{6(2I+1)} \NAtoms.
\label{eq:variance}
\end{equation}
Here $\NAtoms$ is the number of atoms contributing to the spectrum and $f^{\alpha}$ is the single-atom hyperfine spin value \cite{PhysRevA.106.023112}. In this way, it is possible to compute fluctuation spectra for the distinct hyperfine ($\alpha$) components. As evident from Eqs.~(\ref{eq:XSDE}), (\ref{eq:QQT}), and (\ref{eq:variance}), the fluctuating drive term $Q  d{\cal{\mathbf{\hat{W}}}}(t)$ originates in the discreteness of the atomic spin. The equal-time covariance Eq.~(\ref{eq:variance}) describes a separable state, as befits the mean-field description \cite{PhysRevA.106.023112}.

A Faraday rotation signal from such a medium
has power spectral density \cite{ QncuavSERFootnote}
\begin{equation}
\begin{split}
S_{\rm{at}} (\nu) = &   \mathcal{A} \,
r^2 G^2 P^2 \Big[  g_a^2  S_{\hat{F}_z^{a},\hat{F}_z^{a}} (\nu)+  g_b^2 S_{\hat{F}_z^{b},\hat{F}_z^{b}} (\nu)\\ 
 &- g_ag_b  \left(S_{\hat{F}_z^{a},\hat{F}_z^{b}} (\nu)+S_{\hat{F}_z^{b},\hat{F}_z^{a}}(\nu) \right) \Big ], 
 \label{eq:psd}
 \end{split}
\end{equation}
 where $\mathcal{A}$ is a unitless scale factor and $g_\alpha$ is a detuning-dependent coupling proportional to the vector polarizability for the hyperfine state $\alpha$.

\begin{figure}[t]
	\centering
\includegraphics[width=0.9\columnwidth]{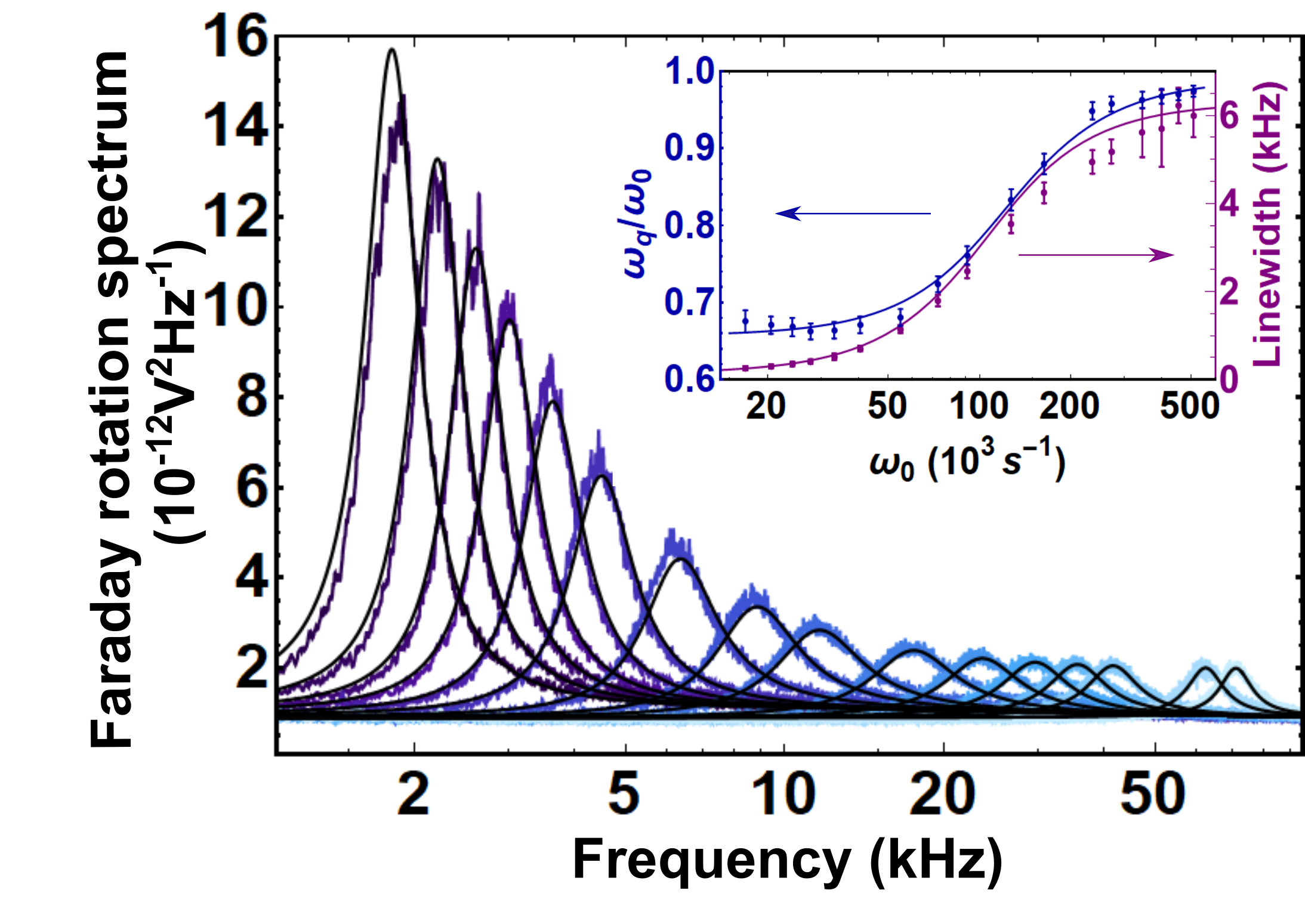}	
	\caption{Single-sided power spectral density (PSD) of the  polarimeter signal (in volts, conversion to rotation angle \SI{35}{\milli\radian/ \volt}) for transverse magnetic fields ranging from \SI{280}{\nano\tesla} to \SI{12}{\micro\tesla} while the vapor cell is maintained at approximately \SI{169}{\celsius}.  Each spectrum shows the linear average \cite{QncuavSERFootnote} of 150 spectra, each computed on a \SI{0.5}{\second} acquisition with a sampling rate of \SI{200}{\kilo\sample\per\second}. A \SI{20} {\hertz} (ten-bin) boxcar smoothing has also been applied \cite{lucivero2017sensitivity}. Black solid lines: fit of Eq.~\eqref{eq:OpticalSpectrum} (excluding 1/f and electronic noise) to the observed spectra (see text). Inset: Left axis shows spin-noise precession frequency $\omega_q$ normalized to $\omega_0= g_s \mu_{B} B/ [\hbar(2I+1)]$, versus $\omega_0$ known by calibration of the coils at low density \cite{QncuavSERFootnote}. Right axis shows the spin-noise linewidth (HWHM) versus $\omega_0$. Data are obtained by fitting the spectra with a distorted Lorentzian (see text). Error bars show $\pm$ one standard deviation in the fit estimation-parameters over 150 acquisitions. Blue (purple) solid line shows $\mathrm{Im}[\lambda]$ ($\mathrm{Re}[\lambda]$) of the eigenvalues of the drift matrix $A$, as given by Eq.7 of \cite{QncuavSERFootnote}. The parameters are discussed in the main text. 
 } 
\label{fig:spec_supp}
\end{figure}

Cross-correlations between the two ground-state hyperfine levels allows for the $g_a g_b$ term in Eq.~\eqref{eq:psd} to partially cancel the $g_a^2$ and $g_b^2$ terms, thereby distorting the spectra and affecting the distribution of spin-noise power.   The non-Lorentzian character of these spectra is illustrated in Fig.~\ref{fig:setup}b) and \ref{fig:setup}c). Representative spin-noise spectra, acquired as a function of transverse bias field, are shown in Fig.~\ref{fig:spec_supp}.

We fit the observed spectra with $S_\mathrm{opt}(\nu) = S_{\rm{at}}(\nu)+ S_{\rm{psn}}$, with $S_{\rm{at}}(\nu)$ from Eq.~\eqref{eq:psd} and photon shot noise $S_{\rm{psn}}=\SI{0.91e-12}{\volt\squared\per\hertz}$ from an independent measurement. The magnetic field is inferred from the current in the $B_x$ coil, previously calibrated by spin-noise spectroscopy at low density \cite{QncuavSERFootnote}. A simultaneous fit to all spectra finds best-fit parameters $\Rse=3.02 \times 10^5$ \si{\per\second}, $\Rsd=0.03 \times 10^5$ \si{\per\second}, $R = 400$ \si{\per\second},  and $\mathcal{A} = 2.3 $. These are respectively the rates of spin-exchange, spin-destruction and spin-depolarizing processes as defined in \cite{QncuavSERFootnote} \footnote{We distinguish between relaxation processes that depolarize only the electron spin and leave the nuclear spin state intact (``spin destruction''), and processes that thermalize both the nuclear and the electron spins (``spin depolarization'') \cite{Happerbook2010, QncuavSERFootnote}.}. 
The fitted spectra are shown as black lines in Fig.~\ref{fig:spec_supp}, and agree well except at the lowest field strengths.
Deviations from Eq.~\eqref{eq:psd} at  low field are expected due to imperfect compensation of remanent fields, the $\Soof(\nu)$ contribution, and diffusion.
A complete model accounting both for spin-exchange and atomic diffusion effects is still missing from the literature, however diffusion alone has been studied in \cite{Diffusion, ShahamPRA2020}. 
From the fitted value of the spin-exchange rate, the $\SI{169}{\celsius}$ temperature of the vapor and the 
$\SI{1.9e-14}{\centi\meter\squared}$  SE cross-section \cite{seltzerthesis}, we infer an alkali number density of $3.4 \times 10^{14}$ atoms/\si{\centi \meter}$^{3}$.

To visualize the ``slowing-down'' of the spin precession and the linewidth reduction, in Fig.~\ref{fig:spec_supp}~(inset) we compare the observed resonance frequency and linewidth from distorted-Lorentzian fits to individual spectra \cite{QncuavSERFootnote} against the predictions of Eq.~\eqref{eq:psd} with the above fit parameters. As described in \cite{QncuavSERFootnote}, the predicted values can be computed from the real and imaginary parts of the eigenvalues of the drift matrix $A$. This extends the results of \cite{PhysRevA.16.1877} to account for spin-destruction and spin depolarizing processes, for any alkali species.

We now study the spectral redistribution of spin-noise power across the transition from the SE-dominated to the SERF regime. 
The total atomic noise power in this state is given by
\begin{equation}
 \int_{0}^{\infty} S_{\rm{at}}(\nu) d\nu = \frac{1}{2}\mathcal{A} r^2 G^2 P^2 [g_a^2 \var(F^a)  + g_b^2  \var(F^b) ], \label{eq:NoisePowerBW}
\end{equation}
where $\var(F^{\alpha})$, $\alpha \in [a,b]$ are given by Eq.~\eqref{eq:variance}. Since our acquisition is limited by a $100$ \si{\kilo\hertz} Nyquist frequency, the experimentally obtained noise is only a portion of Eq.~\eqref{eq:NoisePowerBW}, as discussed in \cite{QncuavSERFootnote}. We stress that the noise in Eq.~\eqref{eq:NoisePowerBW} is independent of the magnetic-resonance parameters and depends only on the number of probed atoms, the probe intensity and detuning,  and the optical linewidth. 

\begin{figure}[t]
	\centering
\includegraphics[width=0.9\columnwidth]{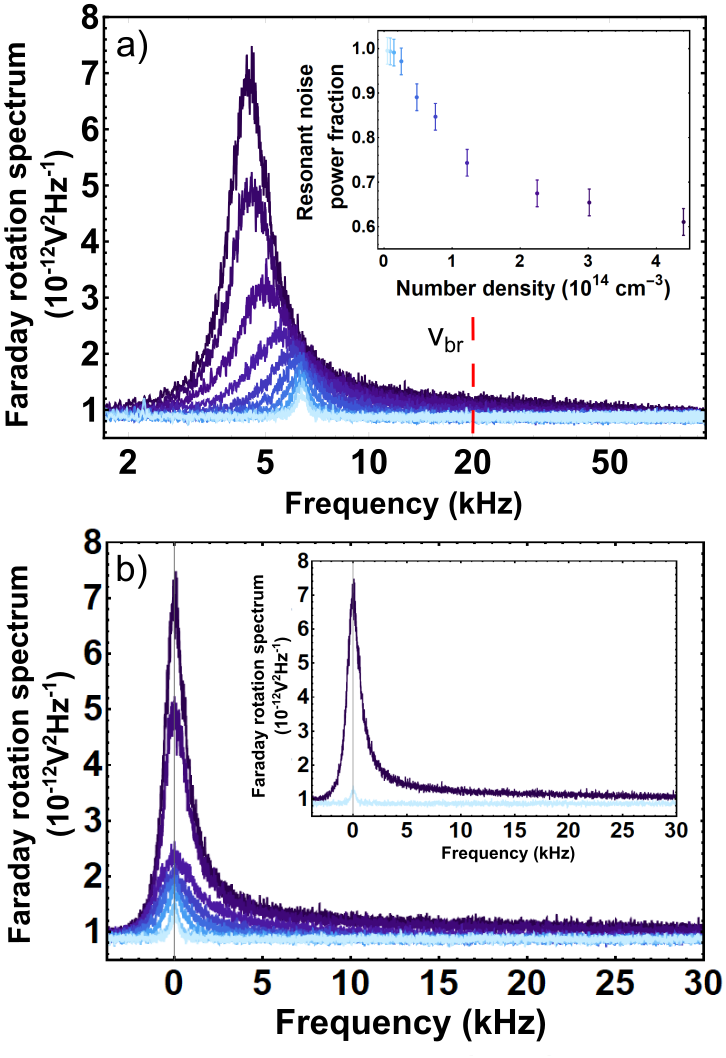}	
\caption{ Non-Lorentzian spectra and apparent (but not real) violation of the area conservation law.
\textbf{a)} Spin-noise spectrum (single-sided PSD) as a function of the \textsuperscript{87}Rb number density for a fixed magnetic field of $B=918 $ \si{\nano\tesla}. Each spectrum shows the linear average of 100 spectra. Long high-frequency tails are apparent. Inset: Resonant noise power fraction as a function of number density as calculated using Eq.~\eqref{eq:N}. The cut-off frequency $\nu_{\mathrm{br}}$ at \SI{20}{\kilo\hertz} is indicated by the red dashed line. Error bars show $\pm$ one standard deviation in the numerical integration over 100 acquisitions. \textbf{b)} For visualizing the power redistribution, 
the spin-noise resonances of Fig.\ref{fig:fig3}a) are plotted with the frequency axis shifted so that each is centred at zero. Curves are plotted for a constant field $B=918 $ \si{\nano\tesla} and varying number density. The values for the number densities are reported in \cite{QncuavSERFootnote}. Inset shows spin-noise resonances for the lower ($5.1 \times 10^{12}$ \si{\per\centi\meter\cubed}) and higher ($4.4 \times 10^{14}$ \si{\per\centi\meter\cubed}) atomic densities acquired.  
 } 
\label{fig:fig3}
\end{figure}

In the SERF regime, the predicted spectra are non-Lorentzian, with a significant portion of spin noise spread over the high-frequency part of the spectrum. To demonstrate this, we acquire spectra under a fixed transverse field of $B=918$ \si{\nano \tesla}, while the alkali number density is varied across the transition from slow SE ($\Rse \ll \omega_0$) to  rapid SE ($\Rse \gg \omega_0$), see Fig.~\ref{fig:fig3}, inset. We numerically integrate the observed spectra to compute 
\begin{equation}
\left. {\int_{\nu_{\mathrm{low}}}^{\nu_{\mathrm{br}} } S_\mathrm{at}(\nu)d\nu }\right/{ \int_{\nu_{\mathrm{low}}}^{\nu_{\mathrm{bw}}} S_\mathrm{at}(\nu)d\nu}, \label{eq:N}
\end{equation}
which describes the fraction of the observed power below a cut-off frequency $\nu_{\mathrm{br}}$.
We choose $\nu_{\mathrm{bw}}=95$ \si{\kilo \hertz} as the upper and $\nu_{\mathrm{low}}=0.5$ \si{\kilo \hertz} as the lower limits of integration in order to avoid distortions in the noise power due to the digitizer's anti-aliasing filter and the 1/f noise, respectively. The cut-off frequency $\nu_{\mathrm{br}}=\SI{20}{\kilo\hertz}$ is chosen to be a few FWHM above resonance so that, were the line Lorentzian, nearly all the spin noise would be below cut-off.  It is seen in Fig.~\ref{fig:fig3} that at low densities nearly all of the atomic noise is below $\nu_{\mathrm{br}}$, whereas at higher densities, in the SERF regime, nearly 50\% of the noise shifts above $\nu_{\mathrm{br}}$ \footnote{In the SE regime, the transverse projections of $\mathbf{F}_a$ and $\mathbf{F}_b$ each precess at the Larmor frequency, whereas in SERF, only the transverse projection of the net spin $\mathbf{F}_a + \mathbf{F}_b$ precesses in this way. This suggests that the \textit{number} of Larmor-precessing degrees of freedom is important in determining the spin-noise power distribution.}. Choice of $\nu_{\mathrm{br}}$ and the contributions of different hyperfine levels are discussed in \cite{QncuavSERFootnote}.

This line reshaping, if not accounted for, can produce systematic errors in calibration, both in estimation of the atom number from the integrated spin noise, and the photon shot noise based on the flat, high-frequency tail of $S_\mathrm{at}(\nu)$. Although we study the spin-unpolarized regime, similar effects can be expected for weakly-polarized ensembles \cite{shah-vasilakis}. Several magnetometry strategies obtain signals from spin precession at $\omega_q$, and would thus benefit from the noise reduction identified here. This noise advantage exists in addition to the well-known coherence-time advantage in the SERF regime \cite{Allred,Kominis2003, SavukovRomalis}.

The observed line reshaping is not specific to SERF or even to spin systems.  Eq.~\eqref{eq:XSDE}, an inhomogeneous linear equation, describes many physical systems. The eigenvectors $\mathbf{v}_i$ and eigenvalues $\lambda_i = i \omega_i - \Gamma_i$ of $A$ describe modes, necessarily Lorentzian, of response to the drive $Q  d{\cal{\mathbf{\hat{W}}}}$. If the noise is aligned to these modes, i.e., if $Q  d{\cal{\mathbf{\hat{W}}}}(t)=\sum_i q_i  \mathbf{v}_i dW_i(t)$, where $dW_i(t)$ are independent Wiener increments and $q_i$ are scalar weights, then each mode fluctuates independently and the spectrum will be a sum of Lorentzian features. If $Q  d{\cal{\mathbf{\hat{W}}}}$ is not so aligned, it produces intra-mode correlations, as in Eq.~\eqref{eq:psd}, and consequent spectral line distortion \cite{QncuavSERFootnote}. In this way, noise spectroscopy probes both the medium's dynamical structure and its noise sources.

In conclusion, we have measured and characterized the spin noise of a thermal \textsuperscript{87}Rb in the transition from the SE-dominated to SERF regimes. We observe anomalous noise lineshapes arising from strong coupling of the ground hyperfine spins in the SERF regime. The line reshaping notably reduces the power in the resonant peak, and produces a  broadband component that imitates photon shot noise. The results validate recent theoretical models, improve the accuracy of thermal-state-based noise calibration for spin squeezing and entanglement generation, and suggest a hyperfine-correlation-induced reduction in fundamental quantum noise for optically-pumped magnetometers operating in the SERF regime.

We thank G. Vasilakis, J. Ko\l{}ody\'{n}ski and V.G. Lucivero for useful discussions. JK and EA acknowledge support from the National Natural Science Foundation of China (NSFC) (Grants No. 12005049, No. 11935012). KM acknowledges support from Grant FJC2021-047840-I funded by MCIN/AEI/ 10.13039/501100011033 and by the European Union ``NextGenerationEU/PRTR.'' MHR acknowledges support from Ayuda PRE2021-098880 financiada por MCIN/AEI/ 10.13039/501100011033 y por el FSE+.  MHR, AS, KM and MWM acknowledge the Spanish Ministry of Science MCIN with funding from NextGenerationEU (PRTR-C17.I1) and by Generalitat de Catalunya, ``Severo Ochoa'' Center of Excellence CEX2019-000910-S; projects  SAPONARIA (PID2021-123813NB-I00) and MARICHAS (PID2021-126059OA-I00) funded by MCIN/ AEI /10.13039/501100011033/ FEDER, EU; Generalitat de Catalunya through the CERCA program;  Ag\`{e}ncia de Gesti\'{o} d'Ajuts Universitaris i de Recerca Grant No. 2017-SGR-1354;   Fundaci\'{o} Privada Cellex; Fundaci\'{o} Mir-Puig; The European Commission project OPMMEG (101099379).

\bibliography{refs}

\end{document}


\preprint{APS/123-QED}

\newcommand{\myaffiliation}{\affiliation}

\newcommand{\ICFO}
{
	\myaffiliation{ICFO - Institut de Ci\`encies Fot\`oniques, The Barcelona Institute of Science and Technology, 08860 Castelldefels (Barcelona), Spain}}

\newcommand{\ICREA}{\myaffiliation{ICREA-Instituci\'{o} Catalana de Recerca i Estudis Avan{\c{c}}ats, 08010 Barcelona, Spain}}

\newcommand{\HDU}
{
	\myaffiliation{Department of Physics, Hangzhou Dianzi University, Hangzhou 310018, China}
}

\newcommand{\FLine}
{\myaffiliation{FieldLine Industries, Boulder, Colorado 80303, USA}}

\newcommand{\Suppl}{Supplementary information}

\author{K. Mouloudakis}
\email{KM and JK contributed equally to this work.}
\ICFO

\author{J. Kong}
\email{KM and JK contributed equally to this work.}
\HDU

\author{A. Sierant}
\ICFO

\author{E. Arkin}
\HDU

\author{M. Hern\'{a}ndez Ruiz}
\ICFO

\author{R. Jim\'{e}nez-Mart\'{i}nez}
\FLine

\author{M. W. Mitchell}
\ICFO
\ICREA

{\onecolumngrid
\centering
\noindent
{\Large {Supplementary Information for: \textit{Anomalous noise spectra in a spin-exchange-relaxation-free alkali-metal vapor}}}
\maketitle
\date{\today}
}
{\onecolumngrid
\tableofcontents
}

\begin{widetext}
\section{Magnetic field calibration}
For calibrating the DC magnetic field experienced by the atoms, we perform spin-noise spectroscopy at $T \approx 100$ \si{\celsius}. By changing the current applied to the X-coil we follow the atomic response by monitoring the central frequency and the linewidth of the spin-noise resonance. The results are summarized in Fig.\ref{fig:Bcalibration}. The precession frequency and the broadening are obtained by fitting the spectrum with a Lorentzian function $S(\nu) = S_0 + S_L \Gamma/ [\Gamma^2 +(\nu-\nu_0)^2]$, in the vicinity of the resonance. 

The flat response of the linewidth as a function of the applied current indicates safe operation away from the SERF regime as well as the absence of significant magnetic field gradients contributing to the spin-relaxation. The residual fields along the $\mathbf{\hat{y}}$ and $\mathbf{\hat{z}}$ directions have been obtained using the ground-state Hanle resonance (see caption of Fig.\ref{fig:Bcalibration}) and compensated accordingly. The compensation measurement is performed at the temperature where the SERF spin-noise data were taken ( $T\approx 169$ \si{\celsius}). We find a field of about $42$ \si{\nano\tesla} and $34$ \si{\nano\tesla} along the $\mathbf{\hat{y}}$ and $\mathbf{\hat{z}}$ directions, respectively. 

\begin{figure}[htp]
	\centering
\includegraphics[width=0.6\columnwidth]{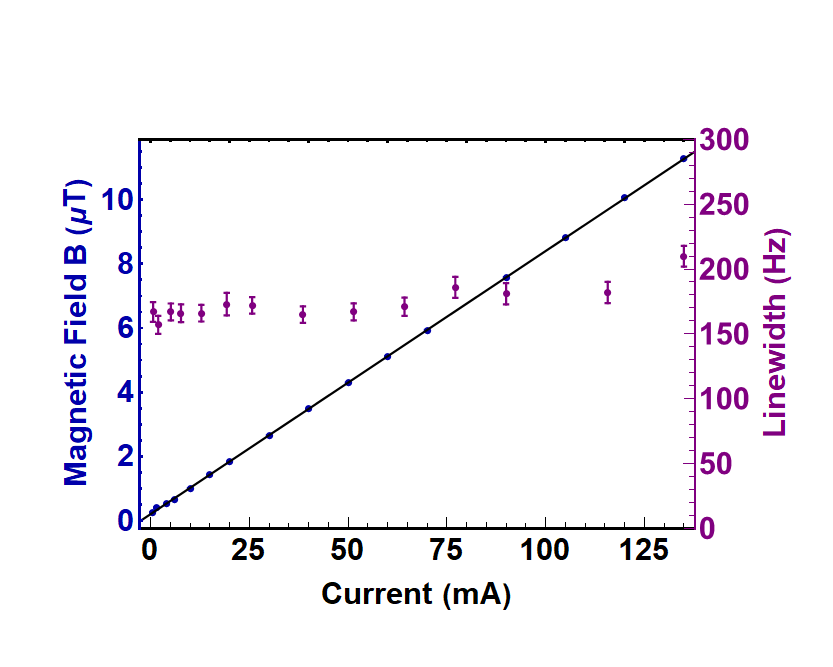}	
\caption{\textbf{Magnetic-field calibration}. Extracted magnetic field and spin-noise linewidth (HWHM) as a function of the current applied to the coil along the $\mathbf{\hat{x}}$-direction. The measurements are performed at a temperature of about $T\approx100$ \si{\celsius}. The spin-noise resonances are fit with a Lorentzian function and the peak frequency is translated into magnetic field units using the \textsuperscript{87}Rb gyromagnetic ratio $\gamma= g_s \mu_B/ [2\pi \hbar (2I+1)] = 6.998$ \si{\hertz\per\nano\tesla}. By fitting the magnetic field versus the applied current (black solid line) we obtain a field to current ratio of $82$~\si{\nano\tesla\per\milli\ampere} and a residual field of $202$~\si{\nano\tesla} along the $\hat{x}$-direction. The calibration equation reads $B(\mu T)=0.202+0.082 I $ where $I$ has units of \si{\milli \ampere}.  At high temperature we have observed an additional DC field originating from the heater with a component along the $\mathbf{\hat{x}}$-direction of $230$~\si{\nano\tesla} that is taken into account. The field components along the other two directions have been measured using the ground-state zero-field resonance of the atoms \cite{CohenTannoudji1970} and have been compensated by applying currents generated by a low-noise low-current source. The errors bars show the standard deviation of 150 fits on the different acquisitions during spectral averaging.} 
	\label{fig:Bcalibration}
\end{figure}
At elevated temperature the vapor enters the SERF regime and both the precession frequency $\omega_q$ and the linewidth of the spin-noise resonance are altered. To experimentally obtain these two features we fit the power spectral density by a sum of a Lorentzian and a dispersive curve: 
\begin{equation}
S(\nu) = S_0 + S_L \frac{\Gamma}{\Gamma^2 +(\nu-\nu_q)^2}  + S_D \frac{\nu-\nu_q}{\Gamma^2 +(\nu-\nu_q)^2}. 
\label{eq:Lor+Dis}
\end{equation}
where $\nu_q =  \omega_q/ 2 \pi$. The fitting results for different magnetic fields are presented in the inset of Fig.2 of the main text. We find that the fitting model of Eq.\eqref{eq:Lor+Dis} accurately captures the resonance features as illustrated with a representative example in Fig.\ref{fig:Lor+Dis_fit}. Eq.\eqref{eq:Lor+Dis} has been chosen in a phenomenological way, however it can be also justified by a more elaborate, first-principle theory for spin-noise in the SERF regime \cite{PhysRevA.106.023112}. 

\begin{figure}[htp]
\centering
\includegraphics[width=0.55\columnwidth]{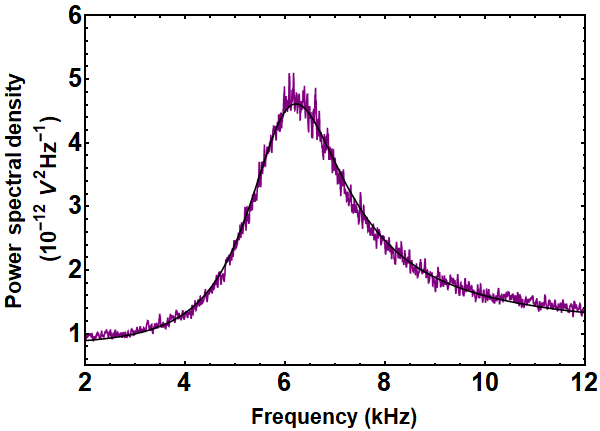}	
\caption{\textbf{SPN peak frequency and linewidth:} A representative fit using Eq.\eqref{eq:Lor+Dis} for finding the peak position and the linewidth of the spin-noise resonance. The fit yields $\nu_q= 5.99 $ \si{\kilo \hertz}. For this particular example, the spectrum is fitted in the vicinity of the resonance in the frequency interval $\nu \in [2.0,12.0]$ \si{\kilo\hertz} such that the resonance is clearly resolved. The corresponding magnetic field is $B \approx 1292$ \si{\nano\tesla}. We note that a simple Lorentzian model is inadequate to capture precisely the peak-position.} 
\label{fig:Lor+Dis_fit}
\end{figure}

\section{Frequency shifts}
In the main text we use the shift of the spin-precession frequency of the spin-noise resonance to estimate the values of the different relaxation rates. Here we justify this reasoning by examining the various frequency shifts that spin-spin and atom-light interactions bring about on the ground-state energy levels of an unpolarized vapor, in the experimentally relevant case of high alkali-metal density and low magnetic fields. In the mean-field approximation, the evolution of the average spin density matrix of a single atom in the ground state is described by the density matrix equation \cite{appelt,ofer}
\begin{equation}
\begin{split}
 \frac{\partial \rho}{\partial t} &= -\frac{i}{\hbar} [A_{\rm{hf}} \mathbf{\hat{I}} \cdot \mathbf{\hat{S}} +g_s \mu_{_B} \mathbf{\hat{S}} \cdot \mathbf{B}+g_I \mu_{_N} \mathbf{\hat{I}} \cdot \mathbf{B}, \rho] -\frac{i}{\hbar} [\delta \nu_{\rm{ls}} + \delta\nu_{\rm{se}}+\delta \nu_{\rm{sd}}, \rho]\\ 
 &+ \Rse [\phi (1+4 \langle \mathbf{\hat{S}} \rangle \cdot \mathbf{\hat{S}})- \rho] 
  + R_{L} [\phi (1+2 \mathbf{s} \cdot \mathbf{\hat{S}})- \rho] + R_{s} (\phi -\rho) +D \mathbf{\nabla}^2 \rho .
  \label{eq:mastereq}
 \end{split}
 \end{equation} 
The equation accounts for the change in the atomic spin state due to the different interactions that take place in the vapor cell. The most prominent are: i) hyperfine interaction between the electron and the nuclear spin, ii) Zeeman interactions of the electron and the nuclear spin with an externally applied magnetic field, iii) binary spin-exchange collisions between pairs of alkali-atoms that conserve the total electron spin of the interacting partners, iv) absorption of light leading to depolarization of the electron's spin, v) spin-rotation interaction which couples the alkali's electron spin to the rotational angular momentum of the colliding pair and vi) wall collisions during diffusion of the atoms into and out of the probe beam in the presence of buffer gas. The effects of magnetic field inhomogeneities on the state evolution are not considered here \cite{PhysRevA.37.2877}. In addition, collisional interactions like SE or interactions that destroy electron's spin polarization (i.e., optical relaxation or $^{87}$Rb -- $^{87}$Rb and $^{87}$Rb -- $\rm{N}_2$ spin-destruction collisions) cause linear frequency shifts accounted by the second commutator of Eq.\eqref{eq:mastereq}. Here, $\delta \nu_{\rm{ls}}$ is the optical light-shift due to the probe beam, $\delta\nu_{\rm{se}}$ is the frequency shift due to spin-exchange collisions and $\delta \nu_{\rm{sd}}$ is the frequency shift due to spin-destruction collisions. The derivation of the master equation and details about each particular interaction can be found in \cite{HapperBook2010,HapperRev}. To elucidate the following discussion on frequency shifts we briefly review each term of Eq.\eqref{eq:mastereq} and we define the various parameters showing up. 

In the density matrix equation, $A_{\rm{hf}}$ is the coupling coefficient for the magnetic dipole interaction between the nuclear spin $\mathbf{\hat{I}}$ and the electron spin $\mathbf{\hat{S}}$, $g_s \approx 2.0$ is the electron g-factor, $\mu_{_B} \approx 9.27 \times 10^{-24}$ \si{\joule\per\tesla} is the Bohr magneton, $g_I \approx 5.5$  is the proton g-factor and $\mu_{_N} \approx 5.05 \times 10^{-27}$ \si{\joule\per\tesla} is the nuclear magneton. At low magnetic fields, the effect of the nuclear magnetic moment is to introduce a negligible frequency shift on the linear precession frequency of the atom by shifting the magnetic sublevels of the two hyperfine manifolds in opposite directions. We will neglect this shift since it is only a tiny fraction of $\omega_0$.  

One way to describe the atom-light interaction is through a polarizability Hamiltonian that couples the light-polarization to the ground-state structure of the atom \cite{PhysRev.163.12}.  For far-detuned linearly polarized light ($\mathbf{s}=0$), the mean light-shift $\delta \nu_{\rm{ls}}$ is a consequence of the rank-1 vector polarizability operator and it is equal to zero. However, fluctuations in the right- and left- circular polarization can introduce light-shift (or backaction) noise, coupling to the spins as an effective fluctuating magnetic field with a white spectrum. It has been shown that light-shift noise has a negligible effect on unpolarized atoms \cite{PhysRevA.106.023112}. More extensive discussions of the effects of light-shift back-action on unpolarized states can be found in \cite{toth2010singlet, BehboodPRL2013, BehboodPRL2014, MitchellNatureCom}. The effect of the rank-2 tensor light-shift is related to the macroscopic feature of linear dichroism exhibited in the medium and is vanishing once the excited-state hyperfine manifolds cannot be resolved \cite{HAPPER197051}. We note that in species different than alkali atoms, like for example in metastable helium, there are significant light-shifts observed through spin-noise spectroscopy  \cite{Heliumlightshift}.

For an unpolarized atom in equilibrium, the expectation value of the electron spin operator is $\langle \mathbf{\hat{S}} \rangle \approx 0$. Therefore, the SE frequency shift $\delta\nu_{\rm{se}}=\Rse\kappa_{\rm{se}} \langle \mathbf{\hat{S}} \rangle \cdot \mathbf{\hat{S}} $, is considered negligible \cite{PhysRevA.16.1877}. In contrast to spin-exchange collisions, the spin-rotation interaction is characterized by collisions of the electron spin with a large rotational angular momentum that can be approximated by a classical vector $\mathbf{N}$. The frequency shift due to the spin-rotation interaction is given by  $\delta \nu_{\rm{sd}}=\langle \bm{\varphi}\cdot \mathbf{\hat{S}}\rangle$ \cite{primer}, where the angle $\bm{\varphi}$ is a vector of magnitude of the order of $10^{-4}$ radians, pointing along the direction of the classical angular momentum vector $\mathbf{N}$, $\bm{\varphi} \mysslash \mathbf{N}$. The brackets $\langle \cdot \rangle$ denote an average over all possible collisional rotation angles. The rotation angle is equally likely to point in any direction, therefore we expect the frequency shift to be negligible.
The term proportional to $\Rse$ describes the spin-exchange interaction, where $\phi = \sum_{i=0}^3 \hat{S}_i \rho \hat{S}_i$ is the electron-spin-depolarized density operator. Here, $\hat{S}_0 \equiv \mathbb{1}/2$ and $\{\hat{S}_1,\hat{S}_2,\hat{S}_3\} = \{\hat{S}_x,\hat{S}_y,\hat{S}_z\}$. Both the photon absorption from the probe beam and the spin-rotation interaction depolarize the electron spin of the atom in the ground state without affecting the nuclear spin state. Both these interactions can be combined into $\Rsd(\phi-\rho)$ with $\Rsd \equiv R_L + R_{\rm{s}}$. 

Finally, the effect of atomic diffusion on the spin-noise spectrum has been recently studied in \cite{Diffusion,Ofer2020}. For a vapor cell with buffer gas and a cell-diameter much larger than the diameter of a well-collimated Gaussian beam, the effect of diffusion is to depolarize the total atomic spin with a rate $R \propto D/(w(z)/2)^2 $, where $D$ is the diffusion constant of the alkali-metal atom and $w(z)$ is the beam waist. Therefore the atomic dynamics in the case of unpolarized atoms can be summarized by the following formula

\begin{equation}
 \frac{d \rho}{d t} = -\frac{i}{\hbar} [A_{\rm{hf}} \mathbf{\hat{I}} \cdot \mathbf{\hat{S}} +g_s \mu_{_B} \mathbf{\hat{S}} \cdot \mathbf{B}, \rho] 
+  R_{\mathrm{se}} (\rho_{\langle \mathbf{S} \rangle} - \rho) + R_{\rm{sd}} (\rho_{\mathbf{0}} -\rho) + R_{} (\frac{\mathbbm{1}}{d} - \rho), 
  \label{eq:mastereq2}
 \end{equation} 
 where,
 \begin{equation}
\rho_{\boldsymbol{\sigma}} \equiv \left( \mathbbm{1}/{2} + 2 \boldsymbol{\sigma}  \cdot \mathbf{S}\right) \otimes \mathrm{Tr}_\mathrm{S}[\rho],
\end{equation}
is the spin-nuclear product state formed by ``resetting'' the spin state to have mean spin $\langle \mathbf{S} \rangle = \boldsymbol{\sigma}$ and leaving the nuclear state unchanged \cite{HapperRev}.

Although the frequency shifts related to the spin-spin and atom-light interactions are negligible in the unpolarized state \cite{PhysRevA.103.L010401,PhysRevResearch.1.033017}, we emphasize that the opposite spin precession in the two ground-state manifolds in combination with the spin-exchange and spin-destruction collisions that transfer angular momentum between the two, can generate significant frequency shifts on the Zeeman sublevels \cite{PhysRevA.16.1877,dikopoltsev2022suppressing} that are quantified in the next section. The preceding shifts allow for precise determination of the spin-exchange rate as demonstrated in the main text and discussed in the following section.  

\section{Coupled hyperfine Bloch equations}

The paramagnetic Faraday rotation signal measured by the balanced photodetector is proportional to the collective angular momentum observables probed by the far-detuned and linearly polarized laser beam, and are defined as $\mathbf{\hat{F}}^{\alpha}\equiv\sum_{i=1}^{\NAtoms } \mathbf{\hat{f}}_{i}^{\alpha}$, $\alpha \in \{ a=I+1/2,b=I-1/2 \}$, where $\hat{\mathbf{f}}_i^\alpha$ is the single-atom angular momentum observable of the $i$th atom, $I$ is the nuclear spin and $\NAtoms$ is the total number of atoms contributing to the detected signal. In the case of an unpolarized atomic ensemble it has been shown \cite{PhysRevA.16.1877,katz2015coherent,PhysRevA.103.043116,PhysRevA.106.023112} that by expanding the density matrix equation of Eq.\eqref{eq:mastereq2} in an irreducible, coupled spherical tensor operator basis, the collective angular momenta operators in the two hyperfine manifolds obey the following first order, coupled stochastic differential equations \footnote{Note this is the case only for unpolarized or low-polarized ensembles under certain assumptions. The validity of these equations has been described in \cite{PhysRevA.106.023112}.}
\small
\begin{eqnarray}
d  \mathbf{\hat{F}^a} (t) &=&  dt \Big\{ \mathbf{\hat{F}^a}(t)  \times \gamma_0 \mathbf{B} -R_{\mathrm{se}} \Big[ \kappa_{aa}  \mathbf{\hat{F}^a}(t)  + \kappa_{ab}  \mathbf{\hat{F}^b}(t) \Big] -\Rsd \Big[ \mu_{aa}  \mathbf{\hat{F}^a} (t) + \mu_{ab}  \mathbf{\hat{F}^b}(t) \Big ]-R \mathbf{\hat{F}^a} (t) \Big\}+q_{aa} d\mathbf{\hat{W}^a}(t)+q_{ab} d\mathbf{\hat{W}^b}(t),
\label{eq:DiffEqA} \nonumber \\
& & \\
d \mathbf{\hat{F}^b}  (t)&=&  dt \Big\{ - \mathbf{\hat{F}^b}(t) \times \gamma_0 \mathbf{B} -R_{\mathrm{se}} \Big[ \kappa_{bb}  \mathbf{\hat{F}^b}(t)  + \kappa_{ba}  \mathbf{\hat{F}^a}(t) \Big]-\Rsd \Big[\mu_{bb}  \mathbf{\hat{F}^b}(t)  + \mu_{ba}  \mathbf{\hat{F}^a}(t) \Big]- R \mathbf{\hat{F}^b} (t) \Big\} +q_{bb} d\mathbf{\hat{W}^b}(t) +q_{ba} d\mathbf{\hat{W}^a}(t).
\label{eq:DiffEqB} \nonumber \\
\end{eqnarray}
\normalsize
Here $\gamma_0=\gamma_e/(2I+1)$ is the atomic gyromagnetic ratio with $\gamma_e = g_s \mu_{_B}/ \hbar \approx 2 \pi \times 2.8 \times 10^4$ \si{\mega\hertz\per\tesla} being the electron gyromagnetic ratio, when neglecting the small contribution of the nuclear Zeeman term, $\omega_0=\gamma_0 B$ the linear Larmor frequency resulting from the diagonalization of the Breit-Rabi Hamiltonian $\hat{H}=A_{\mathrm{hf}} \mathbf{\hat{I}} \cdot \mathbf{\hat{s}} +g_s \mu_{_B} \mathbf{\hat{s}} \cdot \mathbf{B}$ at low magnetic fields (assuming the magnetic field has a component only along a single spatial direction) and $R_{\mathrm{se}}= n \sigma_{\mathrm{se}} u$ is the spin-exchange rate with $\sigma_{\mathrm{se}} \approx 2 \times 10^{-14}$ cm$^2$ being the spin-exchange cross-section and $\bar{u}$ the relative thermal velocity of the pairwise collisions. The coefficients $\kappa_{\alpha\beta}$ and $\mu_{\alpha \beta}$ with $\alpha,\beta \in \{a,b\}$ describe the relaxation and the coupling between the two hyperfine multiplets due to spin-exchange and spin-destruction interactions, respectively. The spin-destructive interactions, like spin-destruction collisions between pairs of alkali atoms or between an alkali atom and an atom of the buffer gas, but also relaxation due to absorption from the probe beam are all captured by the rate $\Rsd \equiv R_s +R_{\mathrm{L}}$. For a $^{87}$Rb vapor with $I=3/2$, the coefficients entering Eqs.\eqref{eq:DiffEqA} and \eqref{eq:DiffEqB} are given by: $\kappa_{aa} =1/8$, $\kappa_{ab} =-5/8$, $\kappa_{ba} =-1/8$, $\kappa_{bb} =5/8$, $\mu_{aa}=7/16$, $\mu_{ab}=-15/16$, $\mu_{ba}=-3/16$ and $\mu_{bb}=11/16$. A detailed analysis of the origin of these coefficients can be found in \cite{PhysRevA.103.043116}. Finally, we note that in this model we ignore effects of the detection backaction on the collective state of the vapor and relaxation caused by the coupling of the alkali spin to the vacuum polarization modes of the probe beam \cite{PhysRevA.70.022106,PhysRevA.90.032113}. 

Regarding the noise terms, $d{\mathcal{\hat{W}}}^{\alpha}_{j} (t)$ with $j \in \{x,y,z\}$ and $\alpha \in \{a,b\}$ represents white Gaussian noise with zero mean and variance $dt$ and are associated with the spontaneous fluctuations of the collective angular momenta in the two hyperfine multiplets. The coefficients $q_{\alpha \beta}$ are the noise-strengths related to the interaction rates $\Rse$, $\Rsd$ and $R$ as a consequence of the fluctuation-dissipation theorem and are such to satisfy the equilibrium variance and covariance values given by $\rm{Tr}[\rho_{\rm{eq}}\hat{F}^{\alpha}_i (0)\hat{F}^{\beta}_j (0)]$, calculated at the many-body fully-mixed state, written as a product state $\rho_\mathrm{eq} \equiv \rho_\mathrm{th}^{\otimes \NAtoms}$, where $\rho_\mathrm{th} \equiv \mathbbm{1}/d$ is the single-atom thermal state, and $d = (2a +1) + (2b +1)$ is the dimension of the single-atom space.  

In the Eqs.\eqref{eq:DiffEqA} and \eqref{eq:DiffEqB} for the spin dynamics we distinguish between ``S-damping'' and ``total-spin depolarizing'' processes. The first are rapid with respect to the nuclear spin dynamics and affect primarily the electron spin while leaving the nuclear spin intact. Processes like these are for example alkali-alkali or alkali-buffer gas spin-destruction collisions (see for example \cite{HapperBook2010}). Under the conditions studied in the experiment (pressure broadening larger than the excited state hyperfine splitting), relaxation from the probe light falls also in this category, resulting in power broadening of the spin-noise resonance. The $\rm{N}_2$ pressure quickly quenches the excited state before the hyperfine coupling has sufficient time to modify the nuclear spin. Wall collisions on the other hand ``thermalize'' both the electron and the nuclear spin and therefore affect the atomic spin state in a different way. It is apparent from Eqs.\eqref{eq:DiffEqA} and \eqref{eq:DiffEqB} that these processes affect differently the hyperfine spin $F$ compared to the ``S-damping'' processes.

The system of differential equations of the two hyperfine angular momenta can be conveniently written in the compact form
\begin{equation}
d \mathbf{\hat{X}} (t) = A \mathbf{\hat{X}}(t) dt+ Q  d{\cal{\mathbf{\hat{W}}}}(t),
\end{equation}
where we define the angular momentum vector $\mathbf{\hat{X}}(t)\equiv[\hat{F}^{a}_x(t),\hat{F}^{a}_y(t),\hat{F}^{a}_z(t),\hat{F}^{b}_x(t),\hat{F}^{b}_y(t),\hat{F}^{b}_z(t)]^T$, the corresponding noise vector 
$d {\cal{\mathbf{\hat{W}}}}(t)\equiv[d\mathcal{\hat{W}}^{a}_x(t),\ldots,d\mathcal{\hat{W}}^{b}_z(t)]^T$, the $6 \times 6$ dynamics matrix $A$ and the noise-strength matrix $Q$.

The real and imaginary parts of the eigenvalues of the dynamics matrix $A$ correspond to the relaxation rate and the precession frequency of the magnetic resonance, respectively. In \cite{PhysRevA.16.1877}, analytical formulas for the eigenvalues as a function of $[I]=2I+1$ were presented, when only spin-exchange collisions were considered as a relaxation mechanism. Here, we derive generalized formulas for the eigenvalues taking into account spin-exchange, spin-destruction and spin-depolarizing collisions. In the high-density and low-field regime, the SE rate exceeds by far the other rates in the system. It is therefore apparent that $\Rsd$ is causing only small corrections in the frequency shift presented in \cite{PhysRevA.16.1877}. 
\begin{equation}
\begin{split}
 \lambda &= - \Big[ R+ \frac{[I]^2+2}{2[I]^2}\Rsd + \frac{[I]^2+2}{3[I]^2}\Rse\pm \sqrt{-\omega_0^2 + \Big(\frac{[I]^2-2}{2[I]^2}\Rsd + \frac{[I]^2+2}{3[I]^2}\Rse \Big )^2 -\frac{2[I]^2-8}{3[I]^4}\Rsd \Rse \pm \frac{i(\Rsd +2\Rse)\omega_0}{[I]}}\Big] \\
 & \approx - \Big[ R+ \frac{[I]^2+2}{3[I]^2}\Rse\pm \sqrt{-\omega_0^2 \pm \frac{2i\Rse\omega_0}{[I]} + \Big( \frac{[I]^2+2}{3[I]^2}\Rse \Big )^2  }\Big] - \mathcal{O} [\Rsd].
\end{split}
\label{eq:eigenvalues}
\end{equation}
\begin{figure}[htp]
	\centering
\includegraphics[width=0.6\columnwidth]{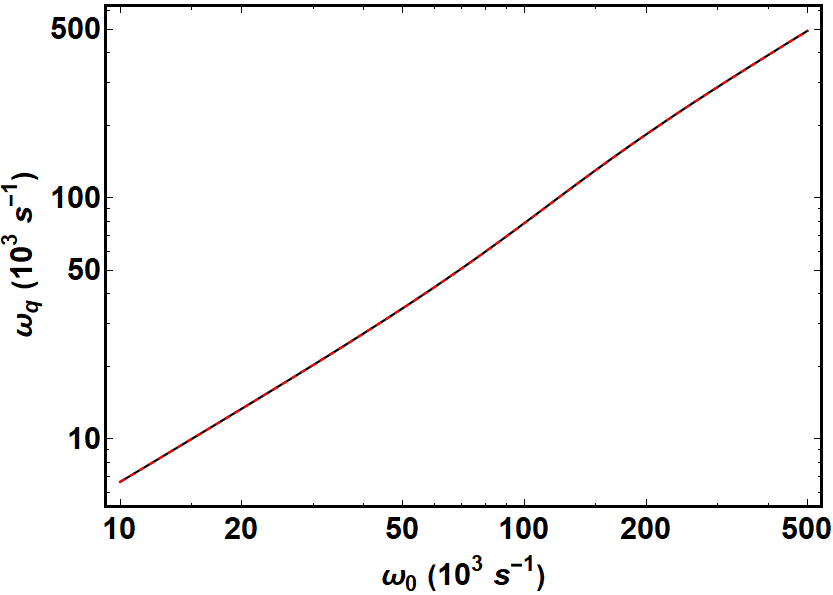}	
	\caption{\textbf{Frequency Shift:} Black solid line: Plot of $\text{Im}[\lambda]$ for $\Rse=3.02 \times 10^5$ \si{\per\second} and $\Rsd=0$. Red dashed line: same calculation with  $\Rse=3.02 \times 10^5$ \si{\per\second} and $\Rsd=0.03 \times 10^5$ \si{\per\second}. } 
	\label{fig:freqshift}
\end{figure}
To illustrate that indeed $\Rsd$ is contributing to only a small fraction in the frequency shift, in Fig.\ref{fig:freqshift} we plot the $\omega_q =\rm{Im}[\lambda]$ as a function of the Larmor frequency $\omega_0$ for zero and non-zero value for $\Rsd$. The red solid line shows the frequency shift only due to SE (zero-order in $\Rsd$), whereas the black dashed line accounts also for the small correction due to spin-destruction. In calculating the transverse eigenvalues we assumed that the 
 DC magnetic field is along the $\mathbf{\hat{x}}-$direction, e.g., $\mathbf{B}= (B,0,0)$. We emphasize that spin-destruction behaves similarly to the spin-exchange collisions, since both interactions affect only the electronic part. As a consequence, in the dynamics for the angular momentum components there are effects similar to the SERF appearing, e.g., frequency shifts and slowing-down of the relaxation time. These effects were recently quantified in \cite{dikopoltsev2022suppressing}. We note spin-depolarizing effects, like wall collisions (diffusion) do not cause any frequency shift, however they can have significant effect on the relaxation rate.

\section{Observables}
In the small angle approximation \cite{julsgaard2001experimental}, the quantum operator associated with the amplified differential photocurrent at the output of the balanced photodetector is proportional to the $\mathcal{\hat{S}}_y (t)$ Stokes parameter, given by \cite{RevModPhys.82.1041,Colangelo2017,colangelo2013quantum}
\begin{equation}
\mathcal{\hat{S}}_y^{\rm{out}} (t) \approx \mathcal{\hat{S}}_y^{\rm{in}} (t) + [g_a \hat{F}_{z}^{a}(t)-g_{b}\hat{F}_{z}^{b} (t)] \mathcal{S}_{x}^{\rm{in}}(t), \label{eq:polarimeter}
\end{equation}
where $\mathbf{\hat{F}^{\alpha}}(t)=(\hat{F}^{\alpha}_x(t),\hat{F}^{\alpha}_y(t),\hat{F}^{\alpha}_z(t))$, with $\alpha \in \{a,b\}$, are the two ground-state angular momentum vector operators. The dimensionless atom-light coupling coefficients $g_{\alpha}$ are defined as \cite{MitchellNatureCom, JimenezMartinezPRL2018}
\begin{equation}
g_{\alpha} \approx \frac{1}{2I+1} \frac{c r_e f_{\mathrm{osc}}}{A_\mathrm{{eff}}} \frac{\nu-\nu_\alpha}{(\nu-\nu_\alpha)^2+(\frac{\Upsilon}{2})^2},
\label{eq:gjDef}
\end{equation}
where $r_e \approx 2.83 \times 10^{-15}$ \si{\meter} is the classical electron radius, $f_{\mathrm{osc}}\approx 0.34$ is the oscillator strength associated with the $\rm{D}_1$ transition of $^{87}$Rb, $2I+1$ is the nuclear spin multiplicity, $A_{\mathrm{eff}}$ is the effective beam area as  defined in the main text, $\Upsilon$ is the full width at half-maximum (FWHM) optical linewidth and $ \nu-\nu_{\alpha}$  the optical detuning. 
The spin-noise power spectral density can be obtained once the following correlation function is calculated \cite{PhysRevA.106.023112}
\begin{equation}
\begin{split}
\mathcal{R}_{\mathcal{\hat{S}}_y^{\rm{out}},\mathcal{\hat{S}}_y^{\rm{out}}}(\tau) & = \frac{1}{2} \langle \mathcal{\hat{S}}_y^{\rm{out}} (\tau) \mathcal{\hat{S}}_y^{\rm{out}}(0)  + \mathcal{\hat{S}}_y^{\rm{out}} (0) \mathcal{\hat{S}}_y^{\rm{out}}(\tau) \rangle  \\
&\approx  \langle \Phi \rangle \delta(\tau)  + \frac{\langle\Phi\rangle^2}{4} \left\{
g_a^2 \mathcal{R}_{\hat{F}_z^a,\hat{F}_z^a}(\tau)  + g_b^2  \mathcal{R}_{\hat{F}_z^b,\hat{F}_z^b}(\tau) - g_a g_b [\mathcal{R}_{\hat{F}_z^a,\hat{F}_z^b}(\tau) + \mathcal{R}_{\hat{F}_z^b,\hat{F}_z^a}(\tau) ] 
\right\}.
\label{Eq:auto-corr}
\end{split}
\end{equation}
with $\langle\Phi\rangle$ being the mean photon flux reaching the balanced polarimeter during a single acquisition run. The first term corresponds to photon shot noise while the second to the auto-correlation of the atomic-noise, separating the contribution of different hyperfine manifolds.

\section{Spin-noise spectrum}
In this section we summarize the main results presented in \cite{PhysRevA.106.023112} and have been adapted in this work to describe and fit the experimental data. To describe the statistical properties of spin-noise it is necessary to use two-time correlation functions: for observables $\hat{A}$ and $\hat{B}$, the covariance is $\mathcal{R}_{\hat{A},\hat{B}}(\tau) \equiv \langle \hat{A}(\tau) \hat{B}(0) + \hat{B}(0)\hat{A}(\tau) \rangle/2= \mathrm{Re}[\langle \hat{A}(\tau) \hat{B}(0) \rangle]$ since $\hat{A}$ and $\hat{B}$ are Hermitian operators and $(\hat{A} \hat{B})^{\dagger}=\hat{B}^{\dagger}\hat{A}^{\dagger}=\hat{B}\hat{A}$. Additionally, given a column vector of observables $\mathbf{\hat{V}}(t) = (\hat{A},\hat{B},\hat{C},...)^T$ with stationary statistical properties, the covariance matrix is 
\begin{equation}
\mathcal{R}_{\mathbf{\hat{V}},\mathbf{\hat{V}}}(\tau) = \frac{1}{2}\Big \{ \langle \mathbf{\hat{V}}(\tau)\mathbf{\hat{V}}^{T}(0)\rangle + [\langle \mathbf{\hat{V}}(0)\mathbf{\hat{V}}^{T}(\tau)\rangle]^{T} \Big \} = \mathrm{Re}[\langle \mathbf{\hat{V}}(\tau)\mathbf{\hat{V}}^{T}(0)\rangle]
\label{eq:RDef}
\end{equation}  
where, $\langle \cdot \rangle $ denotes the average over different and independent realizations of the acquisition process. We assume the equilibrium many-body state as a fully-mixed state, which can be written as the product state $\rho_\mathrm{eq} \equiv \rho_\mathrm{th}^{\otimes \NAtoms}$, where $\rho_\mathrm{th} \equiv \mathbbm{1}/d$ is the single-atom thermal state, and $d = (2a +1) + (2b +1)$ is the dimension of $\rho$. The steady state, equal-time correlators are given by the spin-projection noise, arising from the discreteness of the atomic observables 
\begin{equation}
\mathcal{R}_{\hat{F}_i^{\alpha},\hat{F}_j^{\beta}}(0) = \frac{1}{2} \left\langle \hat{F}_i^{\alpha}(0)\hat{F}_j^{\beta}(0) + 
 \hat{F}_j^{\beta}(0)\hat{F}_i^{\alpha}(0)\right\rangle=
\delta_{ij} \delta_{\alpha \beta} \frac{f^{\alpha}(f^{\alpha}+1)(2f^{\alpha}+1)}{6 (2I+1) }\NAtoms \equiv \delta_{ij} \delta_{\alpha \beta} \var{(F^{\alpha})},
\label{eq:variance}
\end{equation}
where in the last step we defined $\var{(F^{\alpha})} \equiv f^{\alpha} (f^{\alpha}+1) (2f^{\alpha}+1) \NAtoms / [6(2I+1)]$, being the variance of the angular momentum vector operator in the collective unpolarized state.

The drift matrix $A$ and the noise-strength matrix $Q$ are related to the equilibrium covariance matrix $\mathcal{R}_{\mathbf{\hat{X}},\mathbf{\hat{X}}}(0)$ through the self-adjoint relationship \cite{gardiner2009stochastic}
\begin{equation}
A \mathcal{R}_{\mathbf{\hat{X}},\mathbf{\hat{X}}}(0)+\mathcal{R}_{\mathbf{\hat{X}},\mathbf{\hat{X}}}(0) A^{T} =QQ^{T}.
\label{eq:QQT}
\end{equation}
In addition, the $\tau$-dependence of the steady-state covariance matrix is given by
\begin{equation}
\mathcal{R}_{\mathbf{\hat{X}},\mathbf{\hat{X}}}(\tau) =
  \begin{cases}
e^{A\tau}\mathcal{R}_{\mathbf{\hat{X}},\mathbf{\hat{X}}}(0) & ,\tau > 0\\
\\
\mathcal{R}_{\mathbf{\hat{X}},\mathbf{\hat{X}}}(0) e^{-A^{T}\tau} & ,\tau < 0
 \end{cases} \label{eq:cov}
\end{equation}
The power spectral density (assuming infinite acquisition time) is calculated as the Fourier transform of the branched unequal-time correlation function
\begin{equation}
S_{\mathbf{\hat{X}},\mathbf{\hat{X}}}(\omega)  =\int_{-\infty}^{0}  \mathcal{R}_{\mathbf{\hat{X}},\mathbf{\hat{X}}}(0) e^{-A^{T}\tau}  e^{-i\omega \tau} d \tau
+ \int_{0}^{+\infty} e^{A\tau}\mathcal{R}_{\mathbf{\hat{X}},\mathbf{\hat{X}}}(0) e^{-i\omega \tau} d \tau . \label{eq:power spec}
\end{equation}
We can express the power spectral density in terms of the eigenvalues of the matrix $A$ by applying the spectral decomposition $e^{A\tau}=V e^{\Lambda \tau} V^{-1}$ and by explicitly calculating the integral $\int_{0}^{+\infty} e^{[\Lambda-i \omega \mathbb{1}]\tau} d\tau=-(\Lambda-i \omega \mathbb{1})^{-1}$, yielding
\begin{equation}
S_{\mathbf{\hat{X}},\mathbf{\hat{X}}}(\omega)  =  
 -V (\Lambda-i\omega \mathbb{1})^{-1} V^{-1} \mathcal{R}_{\mathbf{\hat{X}},\mathbf{\hat{X}}}(0) + \mathrm{c.c}. \label{eq:power spec1}
\end{equation}
Another useful and equivalent formula for the power spectral density is \cite{gardiner2009stochastic}
\begin{equation}
S_{\mathbf{\hat{X}},\mathbf{\hat{X}}}(\omega)  =  
 -\frac{1}{2 \pi} (A+i \omega \mathbb{1})^{-1} Q Q^T (A^T-i \omega \mathbb{1})^{-1}  \label{eq:power spec2}
\end{equation}
where $QQ^T$ is given by Eq.\eqref{eq:QQT} since the equal-time covariance matrix is known. The preceding formula is preferable for fast numerical computations. Assuming that the probe beam is along the $\mathbf{\hat{z}}-$direction, the total atomic power spectrum would be therefore 
\begin{equation}
 S_{\rm{at}} (\omega) = \frac{r^2 G^2 P^2}{2 \pi} \Big \{ g_a^2  S_{\hat{F}_z^{a},\hat{F}_z^{a}}(\omega)+  g_b^2  S_{\hat{F}_z^{b},\hat{F}_z^{b}}(\omega) - g_ag_b \Big[S_{\hat{F}_z^{a},\hat{F}_z^{b}}(\omega)+S_{\hat{F}_z^{b},\hat{F}_z^{a}}(\omega) \Big] \Big \}.
\label{eq:spectral_components}
\end{equation}
Here $G$ is the transimpedence gain of the PD, $r$ the detector responsivity at $795$ \si{\nano \meter} and $P$ the mean laser power reaching the polarimeter during the acquisition. In Fig.1 of the main text, different spectral components of Eq.\eqref{eq:spectral_components} are plotted as a function of frequency. Integration of the spectra of Eq.\eqref{eq:spectral_components} in the frequency range of the experimentally relevant bandwidth yields the contribution of each spectral component on the total spin-noise power. Each spectrum is subsequently scaled by the off-resonant atom-light coupling coefficients $g_{a,b}$. The integration yields
\begin{equation}
\mathcal{\tilde{P}}_{\rm{at}} =    ~ \mathcal{A} ~ r^2 G^2 P^2 \Big[  g_a^2 \mathcal{P}_{\hat{F}_z^{a},\hat{F}_z^{a}}+  g_b^2 \mathcal{P}_{\hat{F}_z^{b},\hat{F}_z^{b}}  
 - g_ag_b  \left(\mathcal{P}_{\hat{F}_z^{a},\hat{F}_z^{b}} +\mathcal{P}_{\hat{F}_z^{b},\hat{F}_z^{a}}\right) \Big ], 
 \label{eq:spn}
\end{equation}
where $\mathcal{\tilde{P}}_{\rm{at}} = \int_{\nu_\mathrm{low}}^{\nu_\mathrm{high}} S_\mathrm{at} (\nu)d\nu$ with $\nu \in [\nu_\mathrm{low}, \nu_\mathrm{high}] \equiv [\SI{0.5}{\kilo\hertz}, \nu_{\rm{bw}}]$. We limit the integration to \SI{0.5}{\kilo\hertz} in order to avoid low-frequency noise and to the maximum value of \SI{95}{\kilo\hertz} in order to avoid distortion in the noise power due to the anti-aliasing filter (The effect of the anti-aliasing filter of the acquisition system is taken into account in the theoretical calculation and has been simulated using a high-order low-pass filter as discussed in Section ``RMS Spectral averaging'' below).

Finally, we note that besides atomic noise, at low frequencies the excess technical noise can be significant. Sources of this type of noise are mostly thermal fluctuations in the optical fiber, imperfections in the extinction ratio of the Wollaston prism and air-density fluctuations during free-space propagation of the beam. Another source of low frequency noise is the noise spectrum of the longitudinal spin component, centered at $\nu=0$, which can be observed when there is a non-zero field component along the beam direction.

\section{Spin-noise power}

\begin{figure}[t]
	\centering
\includegraphics[width=0.5\columnwidth]{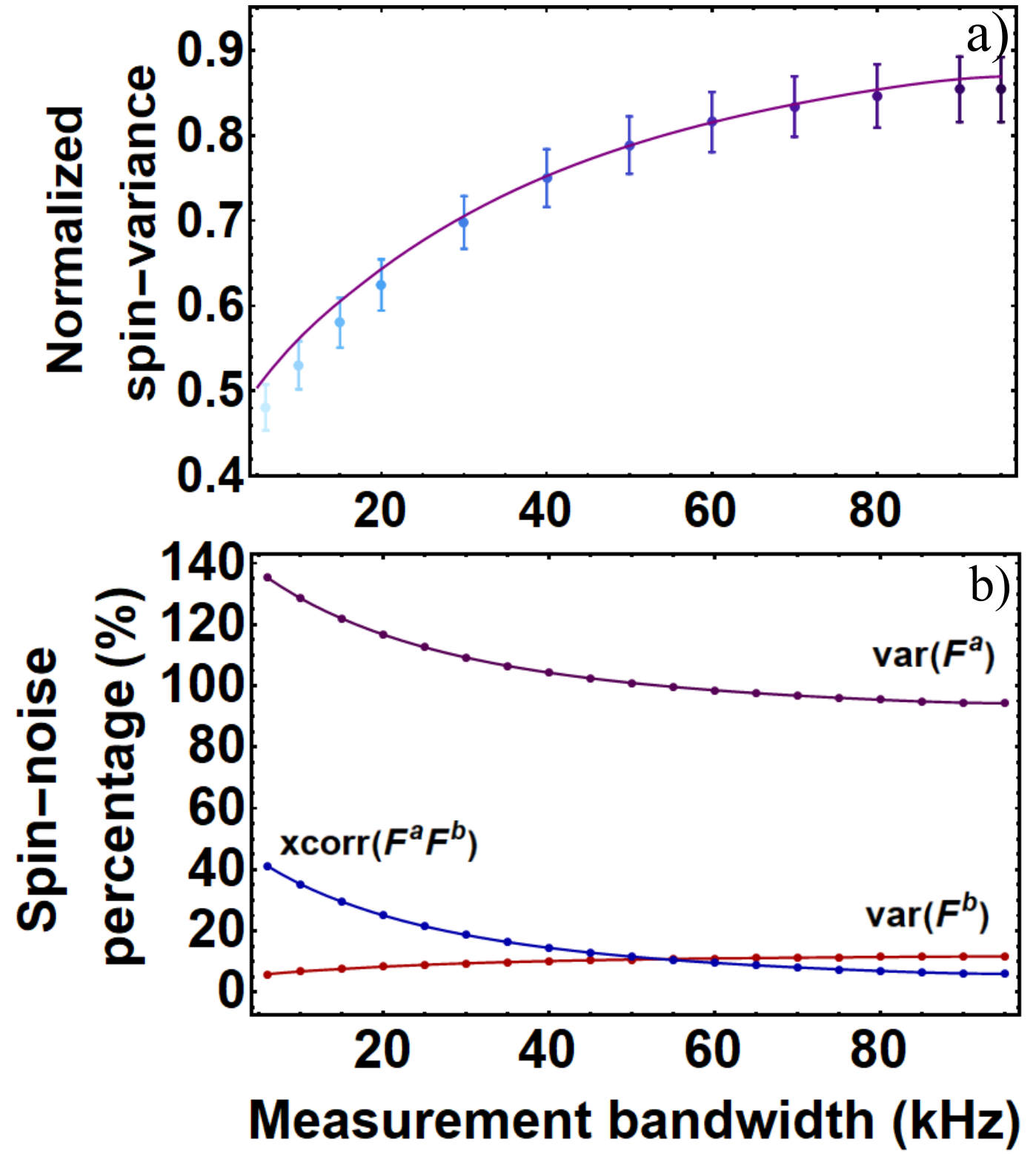}	
	\caption{\textbf{a)} Integrated spin-noise spectrum up to different effective measurement bandwidths $\nu_{\rm{bw}}$ for a fixed magnetic field of $B=385$ \si{\nano\tesla}. The noise is normalized to $g_a^2 \var{(F^a)}+g_b^2 \var{(F^b)}$ (see text). Purple solid line is the prediction of the theory. \textbf{b)} The plot shows theoretical prediction on how the noise obtained in graph \textbf{a)} is distributed at different hyperfine levels, for different measurement bandwidths (see text). Solid lines are interpolations of theoretical predictions indicated by single points in the graph. Purple, red and blue solid lines indicate the noise contribution of $S_{\hat{F}_z^{a},\hat{F}_z^{a}}$ (upper hyperfine manifold), $S_{\hat{F}_z^{b},\hat{F}_z^{b}}$ (lower hyperfine manifold) and $S_{\hat{F}_z^{a},\hat{F}_z^{b}} , S_{\hat{F}_z^{b},\hat{F}_z^{a}}$ (cross-hyperfine) spectra, respectively. } 
	\label{fig:fig3}
\end{figure}

In the SERF regime, because of the non-Lorentzian character of the spectra, a significant portion of spin noise is spread over the high-frequency part of the spectrum. Since the acquisition is limited to 95 \si{\kilo\hertz}, in Fig.\ref{fig:fig3} we show what portion of the total noise is within this frequency range. 
This feature is illustrated in Fig.~\ref{fig:fig3}a) where $S_{\rm{at}} (\nu)$ for $B=385$ \si{\nano\tesla} is numerically integrated up to different $\nu_{\rm{bw}}$ and normalized as  
\begin{equation}
\int_0^{\nu_{\mathrm{bw}} }S_\mathrm{at}(\nu)d\nu / \int_0^\infty S_\mathrm{at}(\nu)d\nu, \label{eq:N}
\end{equation}
where
\begin{equation}
 \int_{0}^{\infty} S_{\rm{at}}(\nu) d\nu = \frac{1}{2}\mathcal{A} ~ r^2 G^2 P^2 [g_a^2 \var(F^a)  + g_b^2  \var(F^b) ], \label{eq:NoisePowerBW}
\end{equation}
is the total atomic noise in the signal for the particular experimental conditions. 

Leveraging the agreement between theory and experiment, in Fig.~\ref{fig:fig3}b) we calculate the way the atomic noise of Fig.~\ref{fig:fig3}a) is distributed among different ground state hyperfine manifolds, for different measurement bandwidths. The noise within the different manifolds is expressed as a percentage of the total noise as obtained in Fig.~\ref{fig:fig3}a). The calculation is based on Eq.\eqref{eq:spn}. It is apparent that the noise in the upper hyperfine manifold can be $120\%$ of the total noise, however strong noise coupling between the two hyperfine levels (up to $40 \%$ of the total noise) is subtracted from the noise in the upper and lower manifolds (see Eq.\eqref{eq:spn}) to obtain the final amount, being the one that is experimentally accessible. We point out that these percentages depend strongly on the detuning, a different choice of which can alter the balance of the noises coupled in the signal \cite{PhysRevA.106.023112}.

\begin{figure}[t]
	\centering
\includegraphics[width=0.5\columnwidth]{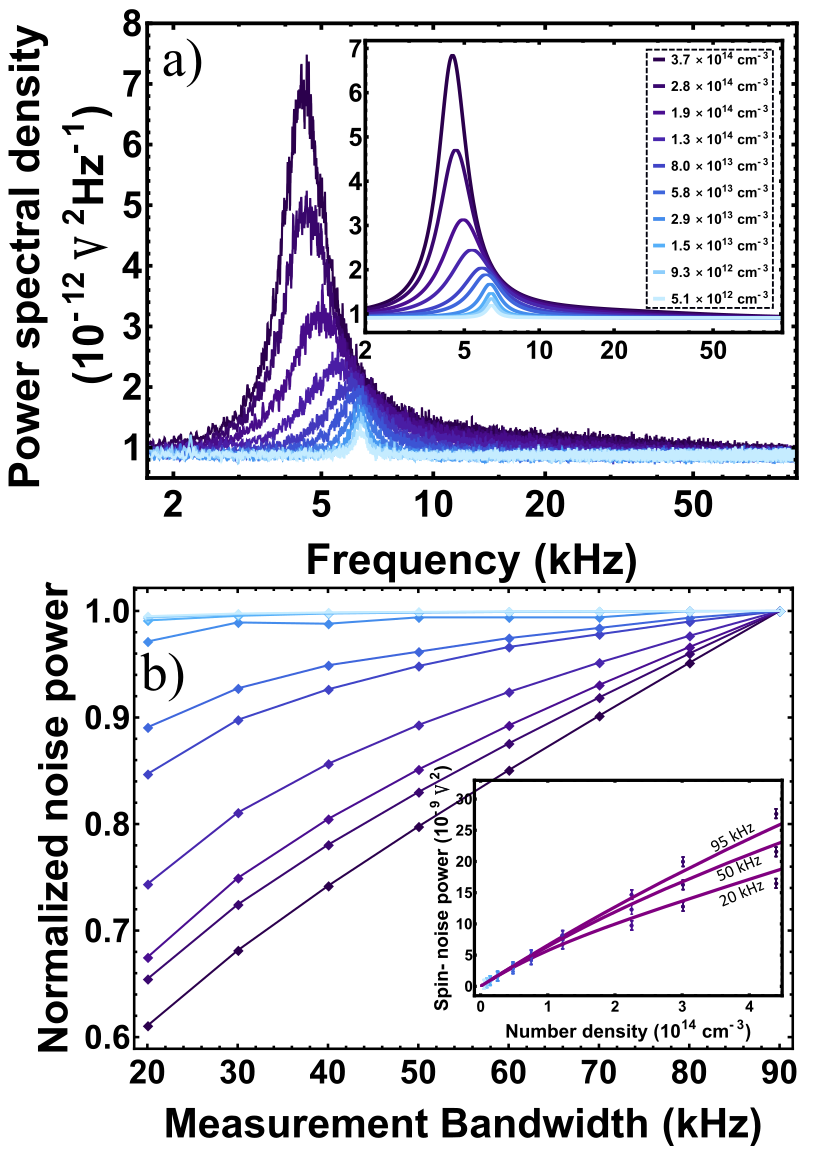}	
	\caption{\textbf{a)} Transition to the SERF regime for a fixed magnetic field of $B=918 $ \si{\nano\tesla} and increasing number density (see legend in the inset), as shown in the inset using the first principles theory. Far wings imitating PSN are apparent in the spectra. \textbf{b)} Spin-noise power as a function of the measurement bandwidth for different number densities, for the spectra presented in \textbf{a)}. The noise is normalized to the noise in the bandwidth of $95$ \si{\kilo\hertz}. Inset shows the scaling of the integrated spin-noise power with the \textsuperscript{87}Rb number density for three different effective measurement bandwidths (20, 50 and 95 \si{\kilo\hertz}). Purple solid lines indicate the theoretical expectation.} 
	\label{fig:fig4}
\end{figure}

In Fig.~\ref{fig:fig4}a), the transition from the slow SE ($\Rse \ll \omega_0 $ ) to the rapid SE regime ( $\Rse \gg \omega_0$) is demonstrated for a fixed magnetic field of $B=918$ \si{\nano \tesla}, while the \textsuperscript{87}Rb number density is altered. The spin-noise power for different cut-off frequencies ( effective measurement bandwidths) as the number density is altered, is shown in Fig.~\ref{fig:fig4}b). In contrast to the slow SE regime where the noise is constant and concentrated in the vicinity of the resonance frequency, there is significant redistribution of noise when the vapor enters the SERF regime that can be controlled as the measurement bandwidth is limited. Finally, in the inset of Fig.~\ref{fig:fig4}b) we show that for different cut-off frequencies the SERF dynamics result in spin-noise power that scales differently with the number density.

\section{RMS spectral averaging}
The spin-noise spectra have been obtained by performing discrete Fourier transform to the acquired time traces $X(j)$ where $j$ is the jth sample. Assuming a total number of samples $N$, we define the FFT as
\begin{equation}
 Y(k)= \frac{1}{N}\sum_{j=1}^{N} X(j) e^{-2 \pi i (j-1)(k-1)/N}   .
\end{equation}
Given $N_{\rm{rep}}$ subsequent acquisitions, the RMS averaging of the complex FFT spectra are obtained as
\begin{equation}
Y_{\rm{rms}}^{R}(k)= \sqrt{\frac{1}{N_{\rm{rep}}}\sum_{i=1}^{N_{\rm{rep}}} \rm{Re}[Y_i(k)]^2} ,
\end{equation}
\begin{equation}
Y_{\rm{rms}}^{I}(k)= \sqrt{\frac{1}{N_{\rm{rep}}}\sum_{i=1}^{N_{\rm{rep}}} \rm{Im}[Y_i(k)]^2}  ,   
\end{equation}
where $Y_i(k)$ is the FFT spectrum of the ith repetition of the acquisition run and $Y_{\rm{rms}}(k) \equiv Y_{\rm{rms}}^{R}(k) +i Y_{\rm{rms}}^{I}(k)$. The power spectral density is therefore
\begin{equation}
S_{\rm{rms}}(k)= \frac{1}{\delta f n_{w}} \Big ( \rm{Re[Y_{\rm{rms}}(k)]^2 + \rm{Im}[Y_{\rm{rms}}(k)]^2} \Big),   
\end{equation}
where $\delta f=1/T_{\rm{acq}}$ is the frequency bin, $T_{\rm{acq}}$ the total acquisition time and  $n_{w}$ is the noise per window bandwidth. We use a uniform window with $n_{w}=1$. We observe that a Hanning window of the form $w(j)=1-\cos{(2 \pi (j-1)/N)}$ is not altering the results presented herein. As mentioned in the main text, in the calculation of spin-noise power, a low-pass filter is applied to simulate the anti-aliasing filter of the DAQ system (see Fig.\ref{fig:Butterworth}).

\begin{figure}[htp]
	\centering
\includegraphics[width=0.5\columnwidth]{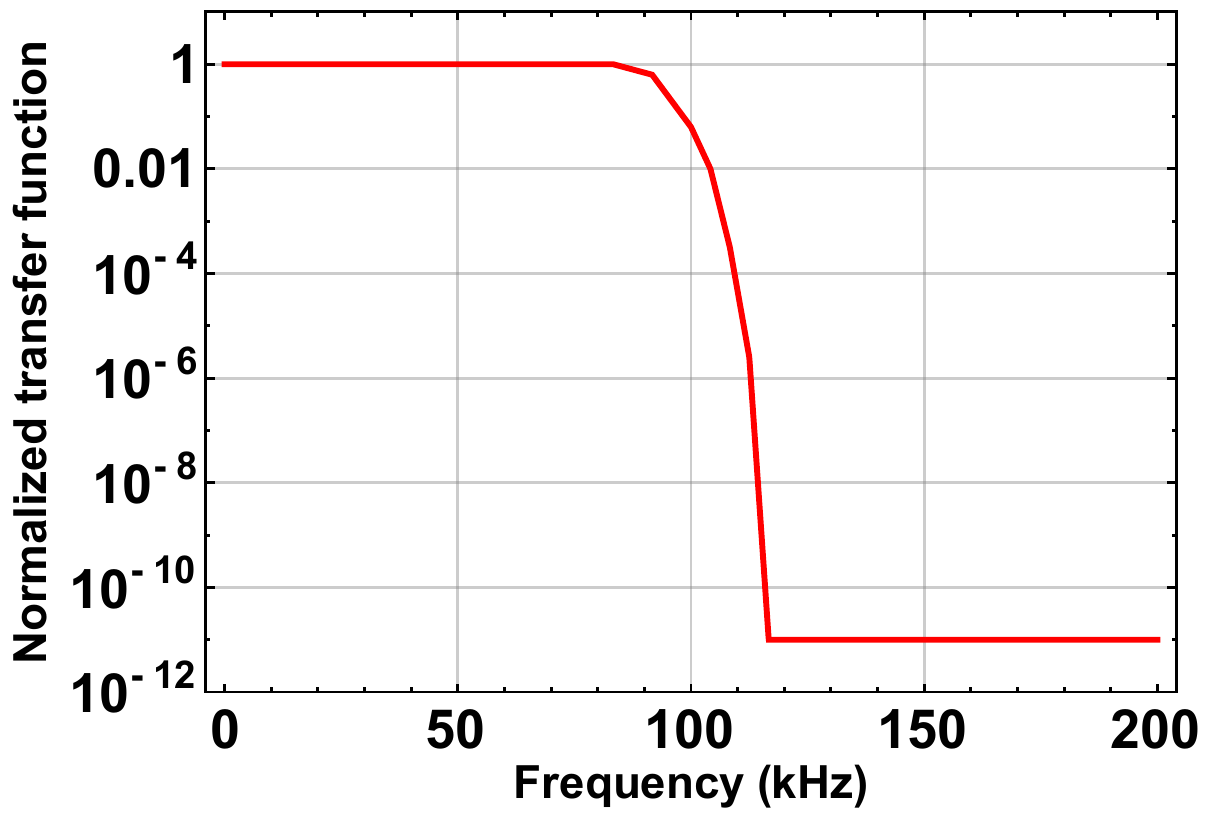}	
	\caption{\textbf{Anti-aliasing, low-pass filter:} To precisely capture the behavior of the anti-aliasing filter applied by the DAQ system on the spin-noise power we simulate the filter using an interpolated function as shown above. The transfer function of the interpolated filter presents a steep roll-off at the cut-off frequency of $96.8$ \si{\kilo\hertz}.  } 
	\label{fig:Butterworth}
\end{figure}

\section{Notes on line reshaping in noise spectroscopy}

The spin dynamics are described by a linear stochastic differential equation for a vector of spin amplitudes $\mathbf{\hat{X}}\equiv[\hat{F}^{a}_x,\hat{F}^{a}_y,\hat{F}^{a}_z,\hat{F}^{b}_x,\hat{F}^{b}_y,\hat{F}^{b}_z]^T$, namely the manuscript's Eq.~(2): 
\begin{equation}
d \mathbf{\hat{X}} (t) = A \mathbf{\hat{X}}(t) dt+ Q  d{\cal{\mathbf{\hat{W}}}}(t).
\end{equation}
The drift matrix $A$ describes relaxation and the coupling of the various components. In the absence of the noise term $Q  d{\cal{\mathbf{\hat{W}}}}(t)$, we would have the homogeneous differential equation 
\begin{equation}
\frac{d}{dt} \mathbf{\hat{X}} (t) - A \mathbf{\hat{X}}(t) = \mathbf{0},
\end{equation}
which can describe the oscillations and eventual relaxation of $\mathbf{X}$ from a given initial condition (It is also used with the regression theorem to calculate correlation functions). The eigenspectrum of $A$ tells us the modes of the system (the eigenvectors will be the coordinates and conjugate momenta) and how they relax and oscillate (the real and imaginary parts of the corresponding eigenvalues).  

In this description of the dynamics, the term $d{\cal{\mathbf{\hat{W}}}}(t)$, which is a vector of independent noise increments, is a forcing term that makes the above system of equations inhomogeneous. For illustration purposes, we write this in terms of a vector of independent, white Langevin noises $\NVector(t)$, rather than the cumbersome but more rigorous vector Wiener increment $d{\cal{\mathbf{\hat{W}}}}(t)$:
\begin{equation}
\left( \frac{d}{dt}  - A \right) \mathbf{\hat{X}}(t) = Q \NVector(t),
\label{eq:PhysicistsStochasticODE}
\end{equation}
As a linear system of equations, the above can be solved by spectral analysis, using the same eigenvectors and eigenvalues of $A$.  In the spectral domain this is solved by
\begin{equation}
\mathbf{\hat{X}}(\omega)= \left(-i \omega \mathbbm{1}  - A \right)^{-1}  Q \NVector(\omega),
\label{eq:PhysicistsStochasticODESolution}
\end{equation}

\begin{figure}[t]
\centering
\includegraphics[width=0.9 \columnwidth]{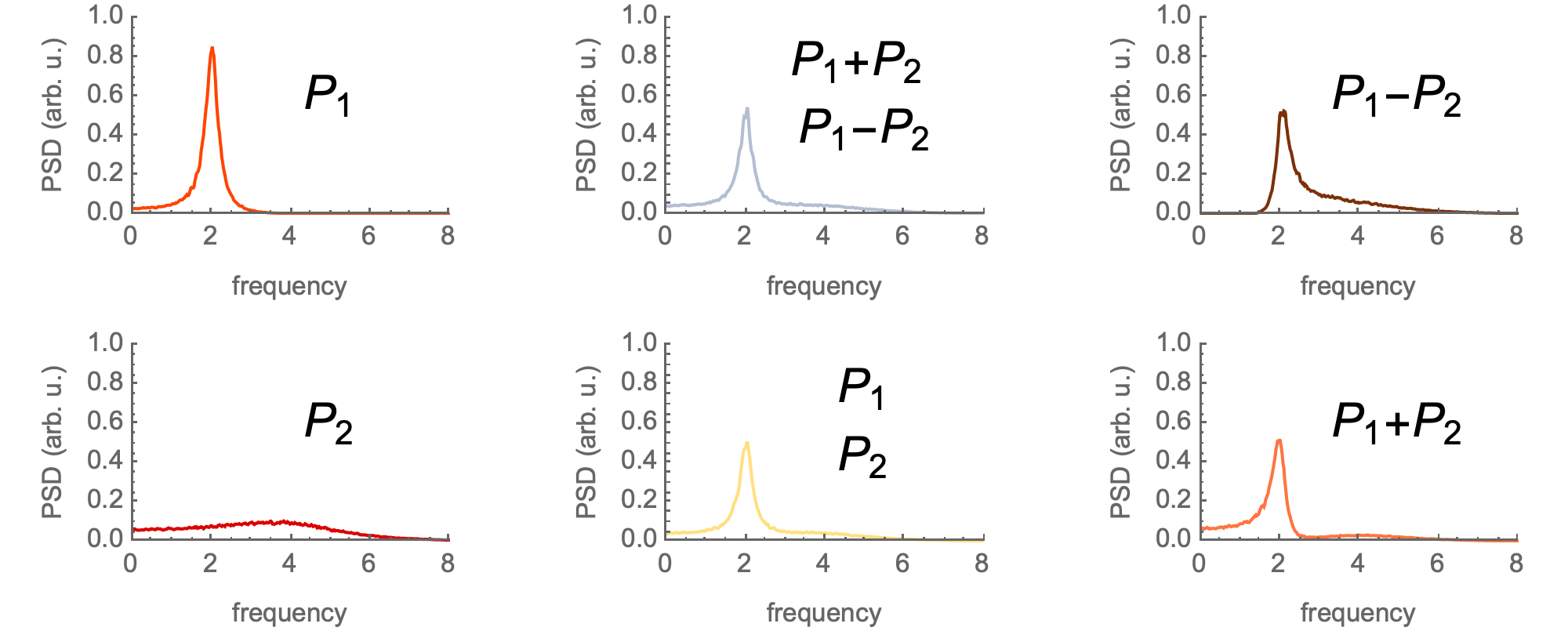}	
\caption{Noise spectra obtained by Monte Carlo simulation of two non-equivalent harmonic oscillators, with coordinates $X_1$ and $X_2$ and corresponding momenta $P_1$ and $P_2$. The $X_1, P_1$ oscillator has resonance frequency 2 and linewidth $\Gamma_1 = 1$, whereas the $X_2, P_2$ oscillator has resonance frequency 4 and linewidth $\Gamma_2 = 4$. Curves show the calculated power spectral density for $(X_1 + X_2)/\sqrt{2}$ for various noise drives, indicated as insets in each sub-figure. Spectra are normalized to have equal area. (Orange, ``$P_1$'') White noise applied  to the $P_1$ variable. This shows the spectrum of the $X_1, P_1$ mode. (Red, ``$P_2$'') White noise applied  to the $P_2$ variable. This shows the spectrum of the $X_2, P_2$ mode. (Yellow, ``$P_1$'' over ``$P_2$'') Independent white noises applied to the $P_1$ and to the $P_2$ variables. More precisely, the applied noise is $\sqrt{\Gamma_1}P_1 \NIndiv_1 + \sqrt{\Gamma_2}P_2 \NIndiv_2$ where $\NIndiv_1$ and $\NIndiv_2$ are independent white noise. The $\sqrt{\Gamma}$ factors compensate for the difference in relaxation rates between the two oscillators, to achieve the same equilibrium variance. We use this same $\sqrt{\Gamma}$ weighting throughout. Note that the resulting spectrum is the (area normalized) sum of the Orange and Red spectra.  (Grey, ``$P_1 + P_2$'' over ``$P_1 - P_2$'') Independent white noises applied to the $(P_1 \pm P_2)/\sqrt{2}$ variables. This should be equal to the Yellow spectrum, and is to within numerical uncertainties.  (Brown, ``$P_1 - P_2$'') White noises applied to the $(P_1 - P_2)/\sqrt{2}$ variable. Relative to the Grey and Yellow spectra, this shows a destructive interference of the $P_1$ and $P_2$ contributions at low frequencies (below the resonance frequency of the $X_1,P_1$ mode, i.e., for frequency $<2$), and a constructive interference at higher frequencies.  (Orange, ``$P_1 + P_2$'') White noises applied to the $(P_1 + P_2)/\sqrt{2}$ variable. Relative to the Grey and Yellow spectra, this shows a constructive interference of the $P_1$ and $P_2$ contributions below the resonance frequency of the $X_1,P_1$ mode, and a destructive interference at higher frequencies. 
}
\label{fig:AsymmetricOscillators}
\end{figure} 

If the driving term $Q \NVector(t)$ is diagonal in the eigenbasis of $A$, i.e., if $Q\NVector(t)$ describes an independent noise for each mode, then the resulting spectrum will simply be a sum of spectra for the different modes. Because the system is linear, each of these contributions will be a Lorentzian with line center and linewidth determined by the imaginary and real parts of the corresponding eigenvalue of $A$.  

At this point we make a parentheses, to show the correspondence to the spin dynamics. In the low-density (or high-field) regime, the hyperfine manifolds $F_a$ and $F_b$ each precess in response to the magnetic field (assumed along the $z$ direction), and slowly relax to zero mean polarization due to infrequent SE collisions. In this regime, the eigenvectors of $A$ are $(F_{ax} \pm i F_{ay})/\sqrt{2}$, $F_{az}$, and $(F_{bx} \pm i F_{by})/\sqrt{2}$, $F_{bz}$. These modes have oscillation frequencies $\pm \omega_L = \pm \gamma B$ for the $x$--$y$ modes, and zero for the $z$ modes.  We note that this describes an oscillation in which $F_{ax}$ and $F_{ay}$ oscillate in the same way as a harmonic oscillator with coordinate and momentum $X$ and $P$. $F_{bx}$ and $F_{by}$ oscillate similarly. 

In contrast, in the SERF regime (high density and low field), the SE collisions plus hyperfine (HF) interactions cause local thermalization of both electron and nuclear spins faster than the spins can precess due to the magnetic field. In this regime, most spin degrees of freedom relax very quickly, at the SE collision rate. An important exception is the total angular momentum (electron spin plus nuclear spin), which is preserved by both SE and HF interactions. This degree of freedom will relax only slowly (due to spin-destruction collisions, wall collisions, and scattering of probe light), while also precessing due to the magnetic field. In the SERF regime, we thus have one high-quality-factor mode  with eigenvectors $(\mathbf{F}_{a} + \mathbf{F}_{b})_x \pm i (\mathbf{F}_{a} + \mathbf{F}_{b})_y$, that oscillates at a frequency related to the Larmor frequency by the slowing-down factor. As above, this describes an oscillation in which $(\mathbf{F}_{a} + \mathbf{F}_{b})_x$ and $(\mathbf{F}_{a} + \mathbf{F}_{b})_y$ oscillate like $X$ and $P$ of a harmonic oscillator. The $z$-polarized total-spin mode $(\mathbf{F}_{a} + \mathbf{F}_{b})_z$ has zero frequency and also decays slowly. Three other eigenvectors, describing other linear combinations of spin components, relax very quickly. 

We believe these observations explain, in a qualitative way
\begin{enumerate}
    \item why the total noise power in Larmor precession peak reduces by about a factor of 1/2 when passing from the low-density to SERF regime. It is because in the low density regime there are two modes (or four eigenvectors) that precess to give finite-frequency narrow features, whereas in the SERF regime there is only one mode (or two eigenvectors). 
    
    \item why the power that is ``lost'' from the Larmor precession peak becomes distributed over a broad bandwidth, and thus mostly at frequencies higher than the Larmor frequency.  It is simply because the SE collisions, which affect everything except the total angular momentum degrees of freedom, are so fast in this regime. 

\end{enumerate}

We now return to discussing \autoref{eq:PhysicistsStochasticODE} and some subtleties that contribute to the shape of the spectrum. We note that the r.h.s., i.e., the driving term $Q \NVector(t)$, need not drive the modes independently, but could drive the different modes in a correlated way. This has consequences for the shape of the noise spectrum, when the noise peaks from different modes overlap. This behavior is, we suspect, the most in need of an intuitive explanation, and also probably the feature that is of greatest interest for other physical systems.

We illustrate with a model that will have analogues in many physical systems. Consider a pair of non-identical harmonic oscillator modes 1 and 2, with coordinates $X_1, X_2$ and corresponding momenta $P_1, P_2$. For this illustration, we assume that mode 1 has a lower oscillation frequency and a higher quality factor than does oscillator 2. These characteristics illustrate well the effects we want to show, and they map onto the SERF system: mode 1 is like the total angular momentum in the SERF system, and mode 2 is like the other transverse spin modes in the SERF system. But the effects we illustrate will be found in any system with multiple modes, and is not restricted to spin systems nor to this particular scenario of frequencies and linewidths. 

If $Q \NVector(t)$ has the form $A_1 P_1 \NIndiv_1(t) + A_2 P_2 \NIndiv_2(t)$, where $A_{1,2}$ are arbitrary amplitudes and $\NIndiv_1(t), \NIndiv_2(t)$ are independent white noise sources, then each mode is independently driven, and the noise spectrum will simply be the sum of the contributions of the two modes. 

In contrast, if the noise does not drive independently the dynamical eigenmodes, for example if $Q \NVector(t) = (P_1 + P_2) \NIndiv(t)$, where  $\NIndiv(t)$ is a single white-noise process, the $1$ and $2$ modes are not driven independently and will rather generate correlations between the $X_1, P_1$ and $X_2, P_2$ oscillations. To illustrate this, we have simulated this scenario (using \textit{Mathematica} functionality for simulating It\^{o} stochastic differential equations), for a variety of drive conditions. See \autoref{fig:AsymmetricOscillators}. Of particular interest are the graphs $P_1 \pm P_2$ (each driven by a single noise term $\frac{1}{\sqrt{2}}(P_1 \pm P_2)\NIndiv(t)$), which show clearly the effect of ``interference'' of the mode 1 and mode 2 contributions, which re-shape the spectrum. 
$P_1 + P_2$ resembles the observed spectra in Figure 3 of the manuscript. 

As described above, this reshaping of the spectra is a general feature of noise spectra, well illustrated in the SERF phenomenon, but certainly not limited to it. It shows that noise spectra contain information not only about the dynamical structure, i.e., oscillation frequencies and relaxation rates, of the system under study, but also the structure of the noise processes that drive this dynamics.

\end{widetext}

\bibliography{refs_SI}